 \documentclass{aastex}

\usepackage{emulateapj5}

\slugcomment{submitted to ApJ on May 8, 2000 ; accepted on Sep 28, 2000}

\shorttitle{UV spectropolarimetry of radio galaxies}
\shortauthors{Kishimoto et al.}

\newcommand{\ha}{H$\alpha$}
\newcommand{\hb}{H$\beta$}
\newcommand{\paa}{Pa$\alpha$}

\begin{document}


\title{UV Spectropolarimetry of Narrow-line Radio
Galaxies\footnotemark[1]}

\author{Makoto Kishimoto and Robert Antonucci}
\affil{Physics Department, University of California, Santa
Barbara, Santa Barbara, CA 93106}

\author{Andrea Cimatti}
\affil{Osservatorio Astrofisico di Arcetri 50125 Firenze, Italy}
 
\author{Todd Hurt}
\affil{Physics Department, University of California, Santa
Barbara, Santa Barbara, CA 93106}

\author{Arjun Dey}
\affil{National Optical Astronomy Observatories, 950
N. Cherry Ave., Tucson, AZ 85719}

\author{Wil van Breugel}
\affil{Institute of Geophysics and Planetary Physics,
Lawrence Livermore National Laboratory, 7000 East Avenue, P.O. Box
808, L-413, Livermore, CA 94550-9900}

\and

\author{Hyron Spinrad}
\affil{Department of Astronomy, University of California at
Berkeley, Berkeley, CA 94720}

\footnotetext[1]{Based on observations with the NASA/ESA Hubble Space
Telescope, obtained at the Space Telescope Science Institute, which is
operated by the Association of Universities for Research in Astronomy,
Inc. under NASA contract No.  NAS5-26555.}

\begin{abstract} 

We present the results of UV spectropolarimetry ($\lambda_{\rm rest}
\simeq 2000 - 3000$\AA) and far-UV spectroscopy ($\lambda_{\rm rest}
\simeq 1500 - 2000$\AA) of two low-redshift narrow-line radio galaxies
(NLRGs) taken with the Faint Object Spectrograph onboard the Hubble
Space Telescope (HST).  Spectropolarimetry of several NLRGs has shown
that, by the presence of broad permitted lines in polarized flux
spectrum, they have hidden quasars seen through scattered light.
Imaging polarimetry has shown that NLRGs including our targets often
have large scattering regions of a few kpc to $\gtrsim 10$ kpc
scale. This has posed a problem about the nature of the scatterers in
these radio galaxies.  Their polarized continuum has the spectral
index similar to or no bluer than that of quasars, which favors
electrons as the dominant scattering particles. The large scattering
region size, however, favors dust scattering, because of its higher
scattering efficiency compared to electrons.

In this paper, we investigate the polarized flux spectrum over a wide
wavelength range, combining our UV data with previous optical/infrared
polarimetry data.  We infer that the scattering would be often caused
by opaque dust clouds in the NLRGs and this would be a part of the
reason for the apparently grey scattering.  In the high-redshift radio
galaxies, these opaque clouds could be the proto-galactic subunits
inferred to be seen in the HST images.  However, we still cannot rule
out the possibility of electron scattering, which could imply the
existence of a large gas mass surrounding these radio galaxies.

\end{abstract}



\section{Introduction}

At least some part of the population of narrow-line radio galaxies
(NLRGs) are believed to harbor quasars in their central region, hidden
from direct view. This has been demonstrated by polarimetric
observations of these radio galaxies.  The observations have shown that
the radiation from some of these NLRGs is linearly polarized with the
E-vector perpendicular to the radio-jet axis. Also the permitted lines
are broad in the polarized flux spectra
\citep{An94,Yo96a,Og97,TCG95,Yo98,Co99}, the same as in Seyfert 2
galaxies \citep{AM85,Tr92,Ka94,Tr95}. In some cases, the broad
components are discernable in total flux, but highly polarized and of
low equivalent width.

The polarization degree often goes higher in the UV than in the
optical due to the lower dilution by the unpolarized star light from
the host galaxies.  Therefore optical, rest-UV polarimetric
observations of high-redshift radio galaxies have been done
extensively \citep{Ci93,di94,di96,Ci96,De96,Ci97,Ci98b,Tr98}. The
imaging polarimetry has shown that the rest-UV image is typically
elongated along the radio axis, and polarized perpendicular to this
elongation \citep{di93,Tr98}. Recently, this has been also
domonstrated by UV imaging polarimetry of nearby radio galaxies
\citep{Hu99}.  Thus, all these facts fit well into the Unified Model
of radio-loud quasars and radio galaxies \citep{Ba89}.  The scattered
light, together with the anisotropic nuclear radiation along the radio
axis, has provided the explanation of at least some part of the
alignment between the UV morphology and radio jet structure in $1
\lesssim z \lesssim 2$ galaxies (called the alignment effect; di
Serego Alighieri et al. 1989; see McCarthy 1993 for a review). In the
two $z\sim4$ galaxies observed, however, the aligned light is
unpolarized starlight (4C41.17, Dey et al. 1997; 6C1909+722, Dey
1998).

While the scattering origin for much of the UV radiation has now been
proven in several NLRGs, the nature of the scattering agents, i.e.
hot/warm electrons and/or dust grains, still remains to be debated.
The imaging polarimetry and long-slit polarimetry have revealed that
the size of the scattering region in some NLRGs is of a few kpc to
$\gtrsim$ 10 kpc scale \citep{Ci96,De96,Tr98,Hu99}. This is one or
more orders of magnitude larger than found in Seyfert galaxies
(e.g. Capetti et al. 1995). This large size implies a rather enormous
mass for the scattering gas in the NLRGs if the scatterers are
electrons \citep{di94,Ci96,De96}. Therefore dust grains are preferred
a priori to be the scattering agent, because of their much better
scattering efficiency (i.e., larger scattering cross section per unit
gas mass).

To address the nature of the scattering regions and environment around
the hidden quasars in the radio galaxies, we need a wide wavelength
coverage for the polarized flux distribution.  In this paper, we
present UV spectropolarimetry of low-redshift narrow-line radio
galaxies obtained by the Hubble Space Telescope (HST). Combining the
data with other optical/IR polarimetric data from the literature, we
seek the general characteristics of the observed scattering regions. 
In \S \ref{sec-obs}, we describe our observation and data reduction
process, and the results are presented in \S \ref{sec-res}. We discuss
the nature of the scattering region in \S \ref{sec-disc}, and
summarize our conclusions in \S \ref{sec-conc}. We adopt $H_0 = 50$
km/sec/Mpc and $q_0 = 0.5$ throughout this paper.

\section{Observation and Data Reduction}\label{sec-obs} 

\begin{figure*}
\plotone{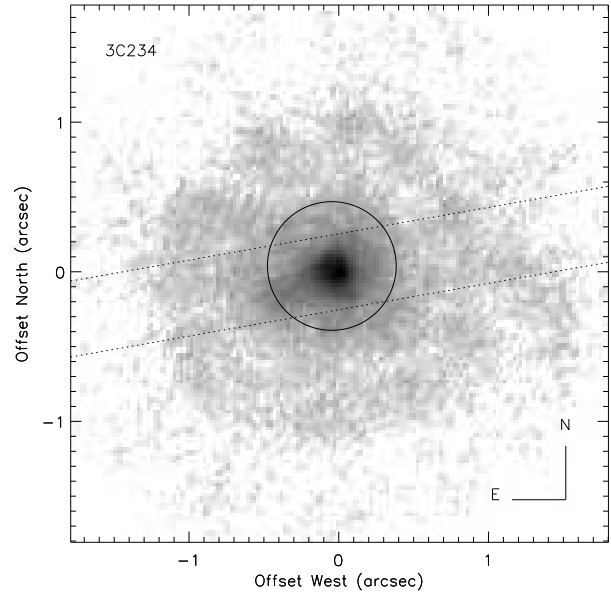} 
\figcaption{The location of the FOS observing aperture ($0.''86$
diameter) for 3C234 is indicated by a circle. The greyscale image is
an $I$ image in log scale from the FOC imaging polarimetry data
\citep{Hu99}. The region between the two dotted lines indicates the
region extracted for calculating the polarized flux distribution in \S
\ref{sec-res}. Note that the FOC image was taken before the COSTAR
installation, while our FOS spectra were taken after the installation.
The image has not been deconvolved.
\label{fig-aper-a}} 
\end{figure*}

Three low-redshift NLRGs 3C234, 3C321 and 3C327 ($z \sim 0.1 - 0.2$)
were observed with the Faint Object Spectrograph (FOS) onboard HST
after the installation of COSTAR. We have used its spectropolarimetry
mode with the grating G270H on the BLUE detector to observe
$\lambda\lambda 2200-3300$\AA\ region at $\sim$ 2.1\AA\ per diode. We
also have taken (non-polarimetric) spectroscopy data with G190H on the
RED detector to observe the far-UV region $\lambda\lambda
1600-2300$\AA\ at $\sim$ 1.4\AA\ per diode.  For these two wavelength
regions, the RED detector has a higher sensitivity than the BLUE
detector, but we used the BLUE detector for the former observation
because of lower geomagnetically induced image motion and lower
instrumental polarization.  The data are summarized in Table
\ref{tab-data}.  The targets were acquired using the 3-stage PEAKUP
procedure of the FOS.  From the target acquistion data, we have
estimated the location of our observing aperture ($0.''86$ diameter)
to the accuracy of $\sim 0.''2$, using the HST/FOC images
\citep{Hu99}. The results are shown in Figures
\ref{fig-aper-a}$\sim$\ref{fig-aper-c}. Unfortunately, we missed the
brightest region for 3C321.

\begin{figure*}
\plotone{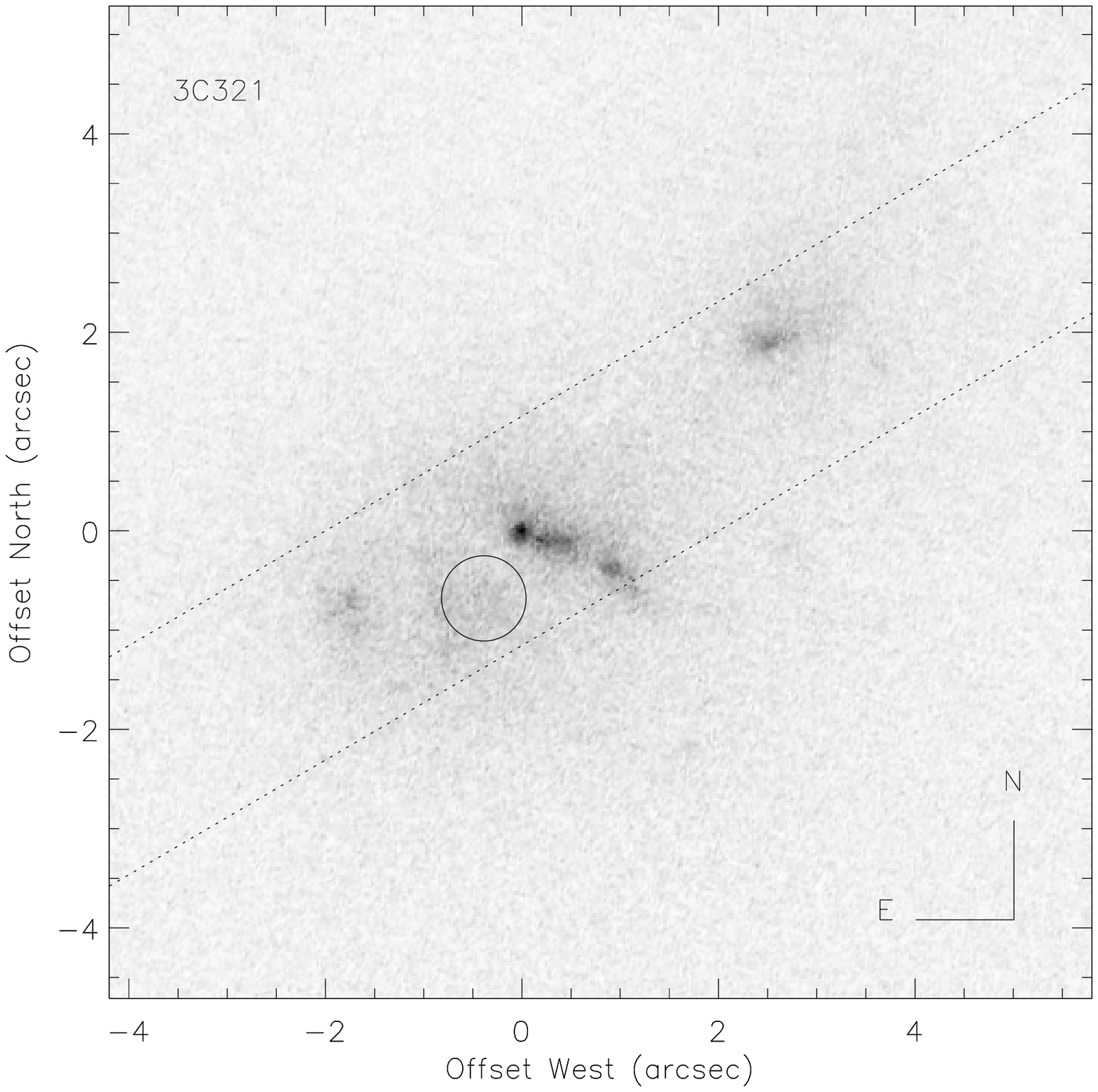}
\figcaption{The same as Fig.\ref{fig-aper-a}, but for 3C321. The image
is in linear scale, and has not been deconvolved. \label{fig-aper-b}}
\end{figure*}

\begin{figure*}
\plotone{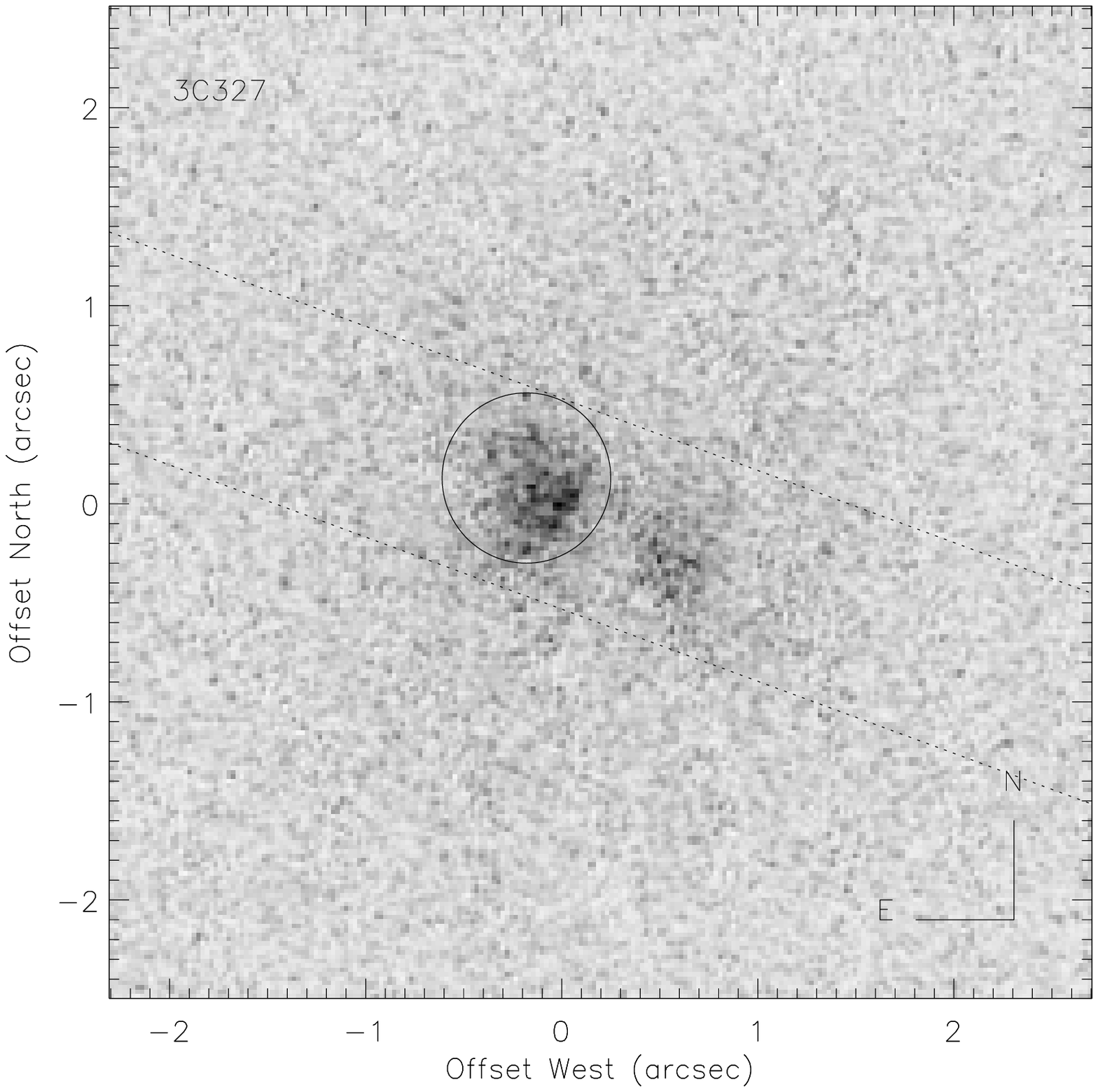}
\figcaption{The same as Fig.\ref{fig-aper-a}, but for 3C327. The image
is in linear scale, and has not been deconvolved. \label{fig-aper-c}}
\end{figure*}

The data were calibrated in a standard way, as described in the HST
Data Handbook (1997). The calibration procedures are (1) correction
for dead diodes, (2) background subtraction, (3) flat-field
correction, (4) wavelength calibration, (5) conversion to absolute
flux. Several pixels with unusual noise are excluded by comparing
several exposures with the same configuration. For the G270H/BLUE
spectra, we have adjusted the zero point of the wavelength
\citep{RKK98}. The heliocentric radial velocity correction was
applied. To obtain the Stokes parameters from the calibrated flux
dataset, we have generated our own script on IDL described briefly
below.  For this process, we did not use the standard software in
STSDAS because of its inadequate handling of the spectropolarimetric
data.

The FOS spectropolarimetry technique is very similar to that in
ground-based instruments. The waveplate is used with a Wollaston
prism, which splits a beam into ordinary and extraordinary rays. These
o-ray and e-ray are observerd almost simultaneously (10 seconds in
turn).  Spectra are taken with the waveplate rotated by 22.\degr5
interval. We obtained our spectra with four different waveplate
positions.  The signal to noise ratio was not good enough to determine
Stokes $V$ parameters with this four-position observation.  This
prevented us from correcting for the instrumental polarization induced
by COSTAR, since this correction needs the circular polarization
information.  However, the effect is expected to be small ($\sim 1$\%
level uncertainty in polarization degree for our wavelength range;
Storrs et al. 1998) compared to our relatively large statistical
errors in polarization. Therefore we calculated the Stokes parameters
$I$, $Q$, and $U$ (after allowing for the wavelength shift between
o-ray and e-ray; Storrs et al. 1998) simply from the relations
\begin{eqnarray}
(1-\cos \delta) Q &=& J^{o}_{0}    - J^{o}_{45} 
                    + J^{e}_{45}   - J^{e}_{0}    \nonumber\\
(1-\cos \delta) U &=& J^{o}_{67.5} - J^{o}_{22.5} 
                    + J^{e}_{22.5} - J^{e}_{67.5} \nonumber\\
I &=& \frac{1}{4} 
          (J^{o}_{0}  + J^{e}_{0} 
         + J^{o}_{45} + J^{e}_{45} \nonumber\\
&& 
     + J^{o}_{22.5} + J^{e}_{22.5} 
     + J^{o}_{67.5} + J^{e}_{67.5}) \label{eq-iqu}
\end{eqnarray}
assuming that the circular polarization is zero (for each intensity
$J$, subscripts denote waveplate positions and superscripts denote
o-ray/e-ray; $\delta$ is the retardation of the waveplate).  We did
not perform any correction for the COSTAR-induced instrumental
polarization. The circular polarization in our objects is expected to
be very small.  After this, the Q and U are corrected for the
variation of waveplate axis over the wavelength and converted from
instrumental coordinates to sky coordinates \citep{AS92}.

We have first subtracted the background using the canonical values
from the standard calibration pipeline.  The FOS background counts
basically correlate with the geomagnetic position of the HST at the
time of the observation \citep{HL96}. The canonical background values
in the pipeline are derived from this information, but there could be
some residual counts.  For the far-UV observation with G190H, there
are unilluminated pixels and we have used them for the scattered light
subtraction and additional background subtraction adjustment. With
G270H (spectropolarimetric data), however, there are no unilluminated
pixels available.  We found that with the canonical background
subtraction, the blue side of the G270H Stokes $I$ spectra did not
match well with the red side of the G190H spectra for each
object. Therefore we scaled the canonical background for G270H spectra
to match with the G190H spectra.  The scaling factors were found to be
1.35, 1.38, and 1.45, for 3C234, 3C321, and 3C327, respectively.
These scaling values are rather large, suggesting that the canonical
backgrounds were systematically underestimated for these observations.
The scattered light in our G270H spectra is expected to be small since
our objects are rather red \citep{Ro94}. We have confirmed the color
of our objects by calculating the optical flux in our observing
aperture using the archival WFPC2/F702W images. The geomagnetic image
motion could affect the spectra, but in our case this effect is
expected to be small because our observing aperture was $0.''86$,
where the height of diodes is $1.''29$ and the typical uncorrected
geomagnetic motion is $0.''15$, and because real-time onboard
corrections were done. Also the telescope pointing was good in our
observations, judging from the available jitter data.

Note that this background scaling, or background itself, essentially
does not affect the unnormalized Stokes parameters $Q$ and $U$
systematically, thus does not affect the polarized flux and position
angle of polarization systematically (but affects $P$, through
$I$). The $Q$ and $U$ are essentially obtained from the difference of
the o-ray and e-ray [see eq.(\ref{eq-iqu})], and these two rays are
observed almost simultaneusly as described above. The FOS background
counts originate mainly from high energy particles hitting the
photocathode or detector itself, and are considered to reside equally
in o-ray and e-ray spectra.

\section{Results}\label{sec-res}

\begin{figure*}
\plotone{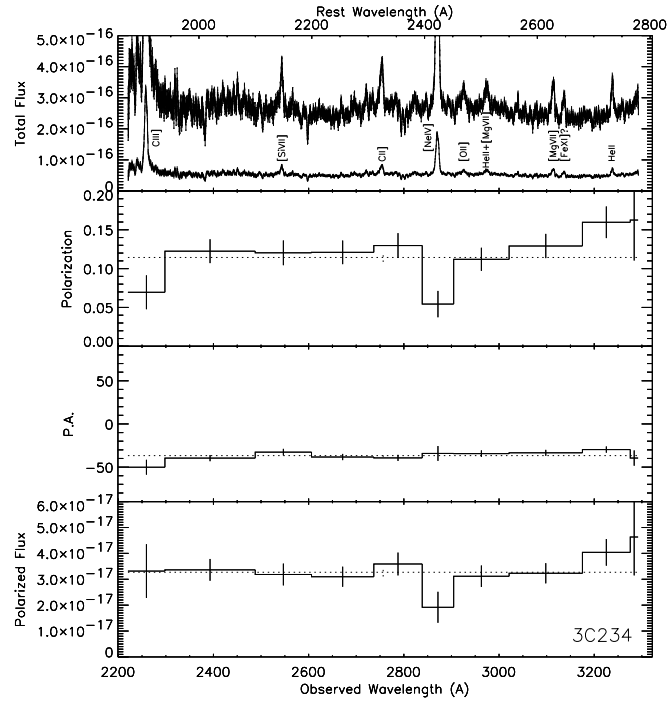}
\figcaption{UV Spectropolarimetry of 3C234.  From top to bottom, $I$
spectra, fractional polarization, position angle, and polarized flux
are shown with 1$\sigma$ error bars.  The flux unit is erg cm$^{-2}$
sec$^{-1}$ \AA$^{-1}$.  The position angle is in degrees.  In the top
panel, two $I$ spectra are shown : the upper one has the true flux,
and the lower one is the true flux multiplied by 1/5.  Both of $I$
spectra have been smoothed by 1 diode width (= 4 pixel).  The line
identifications are shown, and also note the possible bumps from the
\ion{Fe}{2} lines at around $\sim 2400$\AA\ and $\sim 2100$\AA.  In
other panels, the dotted lines show the averaged values over the whole
observed wavelength.  The polarized flux has been debiased following
\citet{SS85}.
\label{fig-pol-3c234}}
\end{figure*}

\begin{figure*}
\plotone{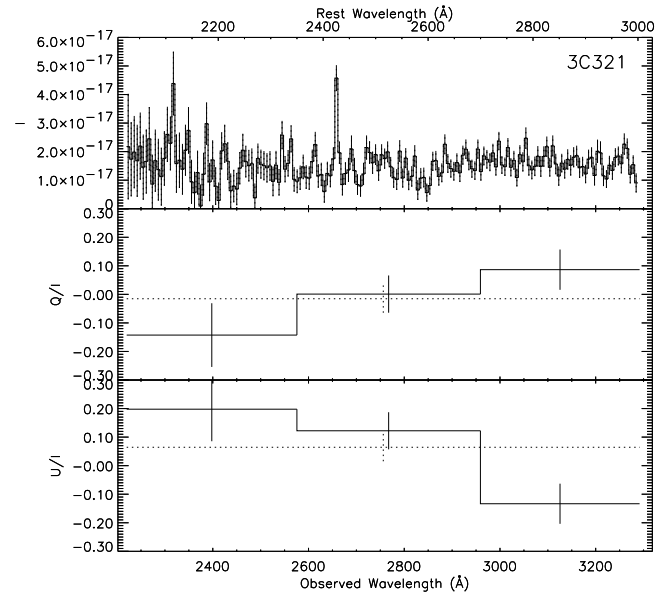}
\figcaption{UV Spectropolarimetry of 3C321. From top to bottom, the
$I$ spectrum (binned with 3 diodes = 12 pixels) and normalized Stokes
spectra $Q/I$ and $U/I$ are shown. Note that the observing aperture
missed the brightest region. \label{fig-pol-3c321}}
\end{figure*}

\begin{figure*}
\plotone{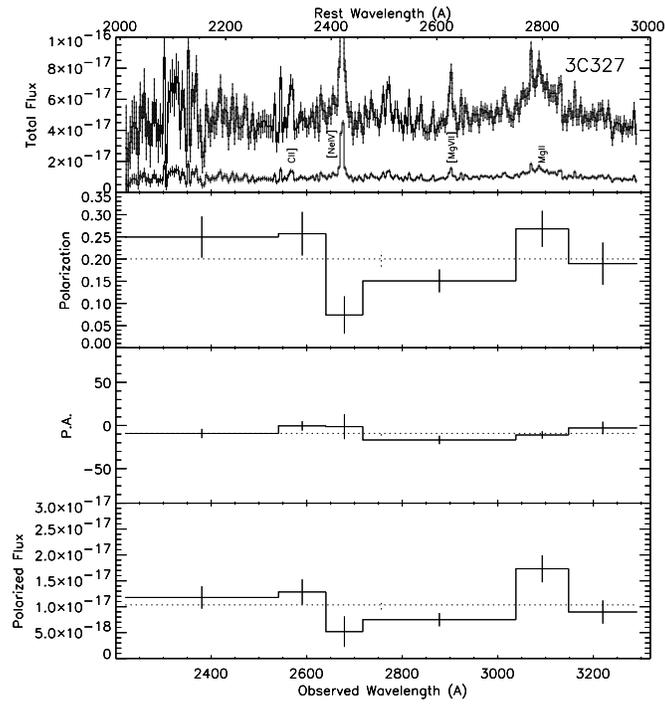}
\figcaption{UV Spectropolarimetry of 3C327.  The styles are the same
as Fig.\ref{fig-pol-3c234}, except that the $I$ spectra are binned
with 2 diodes = 8 pixels. \label{fig-pol-3c327}}
\end{figure*}

\begin{figure*}
\plotone{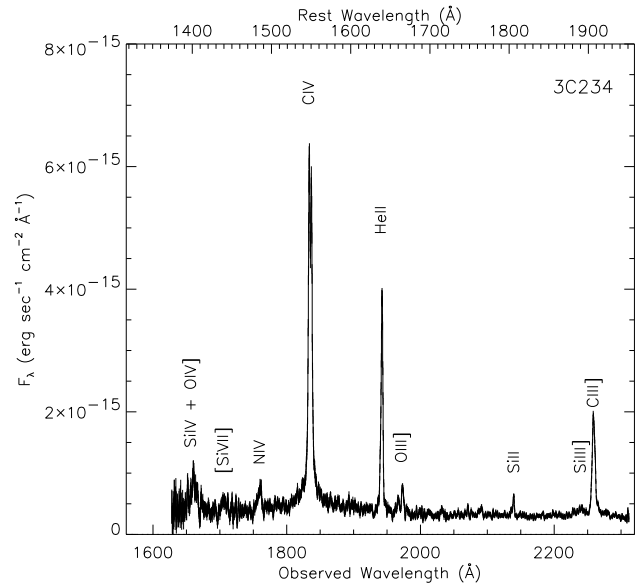}
\figcaption{Far-UV spectrum of 3C234, smoothed by 1 diode (= 4
pixel). \label{fig-fuv-3c234}}
\end{figure*}

Figures \ref{fig-pol-3c234} $\sim$ \ref{fig-pol-3c327} show the
results of our spectropolarimetric observations. The far-UV spectra
are given in Figures \ref{fig-fuv-3c234} $\sim$
\ref{fig-fuv-3c327}. The errors are calculated from the statistical
error of the raw counts.  Due to the aperture miscentering for 3C321,
we could not detect polarization for this object
(Fig.\ref{fig-pol-3c321}).  For 3C234 and 3C327, we have detected high
polarization ($10 \sim 20$ \% level). For these objects, the UV
polarization is roughly constant over the observed UV wavelength
range.

\begin{figure*}
\plotone{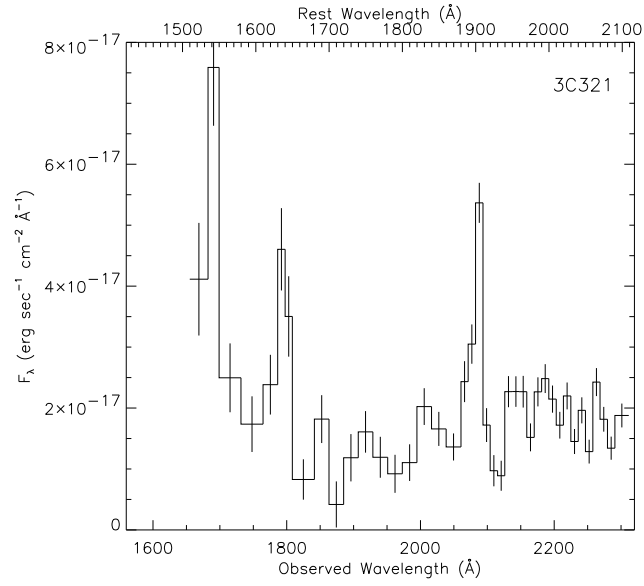} 
\figcaption{Far-UV spectrum of 3C321. Note
that the observing aperture missed the brightest
region.\label{fig-fuv-3c321}}
\end{figure*}

\begin{figure*}
\plotone{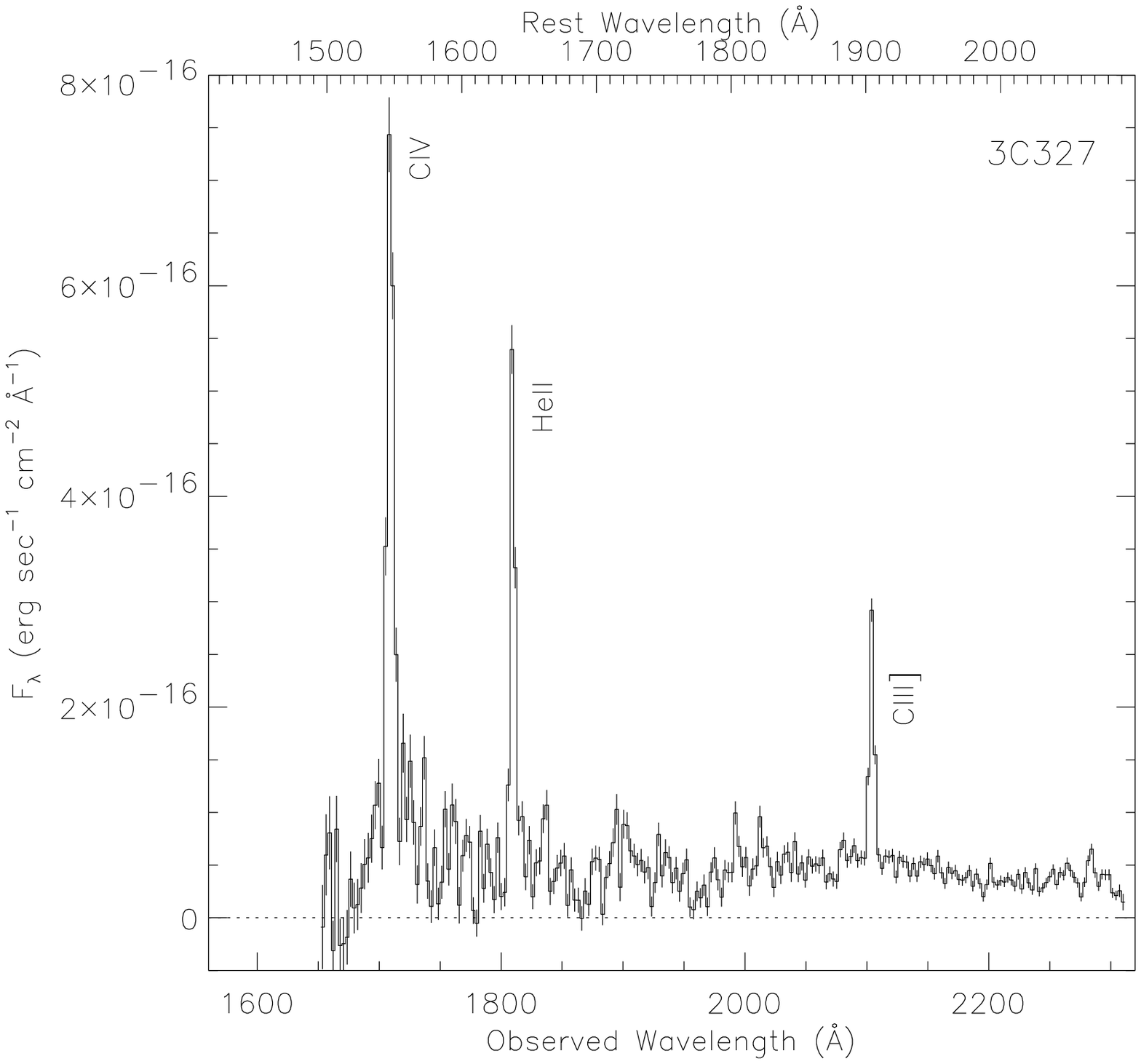} 
\figcaption{Far-UV spectrum of 3C327, binned
with 2 diodes (= 8 pixel). \label{fig-fuv-3c327}}
\end{figure*}

\begin{figure*}
\plotone{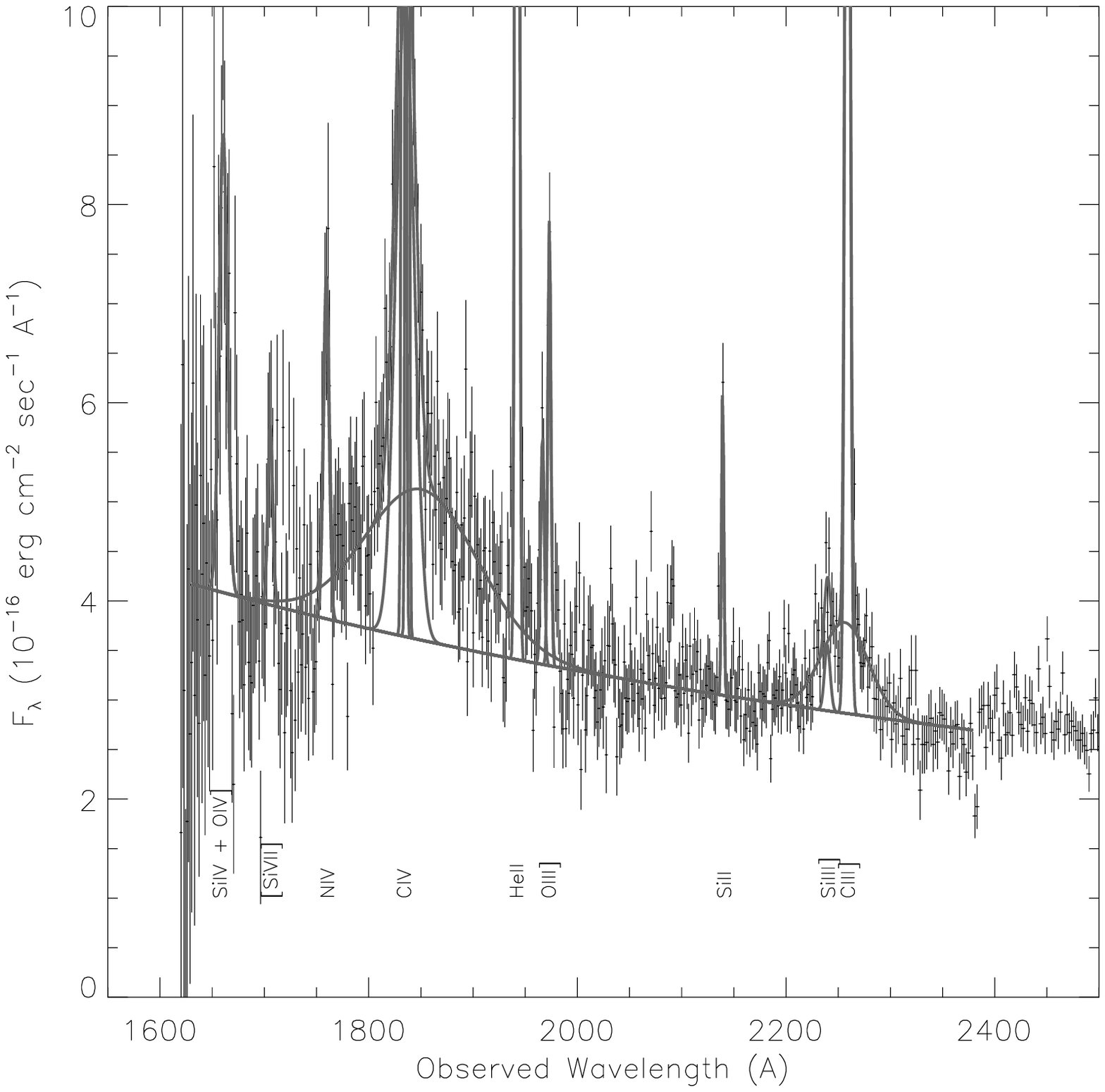}
\figcaption{The line fitting result for the far-UV (G190H) spectra of
3C234. The G270H has been combined at 2280\AA. Table \ref{tab-line}
lists the fitted parameters. The simultaneous fit of the continuum and
several lines indicated was implemented. The obtained continuum slope
is $F_{\nu} \propto \nu^{-0.85\pm0.11}$, but this could be uncertain
due to the low S/N in the short wavelength side.  If the real
continuum is redder as in the longer wavelengh side, the extremely
broad component of the \ion{C}{4} line could be even broader and would
have a larger flux.
\label{fig-line-3c234}}
\end{figure*}

We also show the properties of the emission lines seen in our spectra
in Table \ref{tab-line}, determined by fitting a gaussian for each
line.  For the line widths, the instrumental width of 1 diode has been
subtracted in quadrature. For 3C234, the broad components are seen in
the total flux at \ion{C}{4} $\lambda$1549 and \ion{C}{3}]
$\lambda$1908.  For the former, there seems to be two distinct broad
components with FWHM of $\sim 3000$ km sec$^{-1}$ and $\sim 20000$ km
sec$^{-1}$. We show the line fitting results in Figure
\ref{fig-line-3c234}.  The simultaneous fit of the continuum (power
law) and several lines (gaussians) resulted in having the continuum
slope of $F_{\nu} \propto \nu^{-0.85\pm0.11}$, but this could be
uncertain due to the low S/N in the short wavelength side.  If the
real continuum is redder as in the longer wavelengh side (see
Fig.\ref{fig-pol-3c234}, though note the possible \ion{Fe}{2}
contributions), the extremely broad component of the \ion{C}{4} line
could be even broader and would have a larger flux.  For the broad
component in the \ion{C}{3}] line, there might also be two components,
but the smaller EW and the blend with the \ion{Si}{3}] line prevented
us to do a reliable two-component fit.  For 3C327, we see broad
\ion{Mg}{2} $\lambda$2800 line in the total flux. We tentatively
determined its FWHM to be $\sim 8000$ km sec$^{-1}$, although there
could be contributions from broad \ion{Fe}{2} emission lines in this
region.

As we described in \S \ref{sec-obs}, we have increased the amount of
the background subtraction for the spectropolarimetry data, compared
to the nominal background. The background subtraction affects only
$I$, and does not have an influence on $Q$ and $U$ (unnormalized
Stokes parameters). The object counts are lower at shorter wavelength
which have lower sensitivity, so the fraction of the background counts
are larger there.  Therefore, larger background subtraction results in
smaller $I$ and larger $P$ at the shorter wavelength.  This changes
the total flux level and its spectral shape rather significantly in
3C321 and 3C327 since the background fraction in these objects was
rather high. However, the overall wavelength dependence of $P$ does
not change significantly compared with the relatively large error in
$P$. The polarized flux is not influenced, as the $Q$ and $U$ do not
change.

\begin{figure*}
\plotone{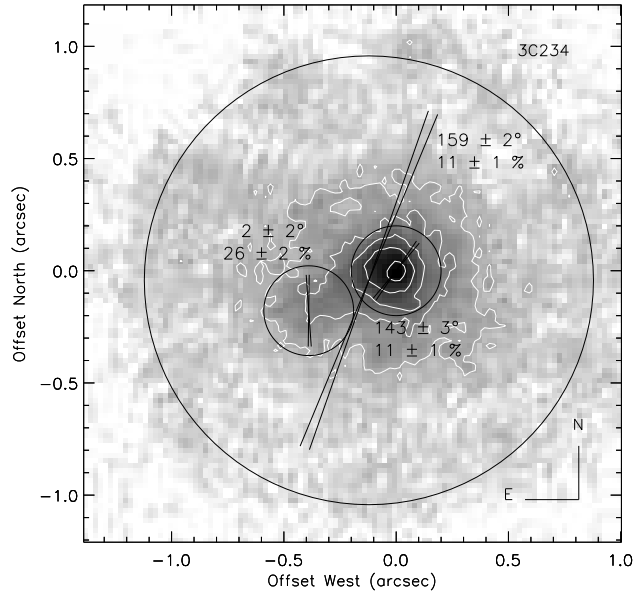}
\figcaption{Imaging polarimetry of 3C234 \citep{Hu99}, with three
different synthetic apertures. The PA and polarization are written
with each aperture, together with two lines indicating the directions
of PA$\pm \sigma_{\rm PA}$. The image is the same as
Fig.\ref{fig-aper-a}. Contours in percent of the peak are 5, 10, 20,
40, 80.  Note that the polarization off the nucleus is only suggestive
since this data set is pre-COSTAR and the influence of the central
bright point-like source is hard to quantify.
\label{fig-imgpol-3c234}}
\end{figure*}

\begin{figure*}
\plotone{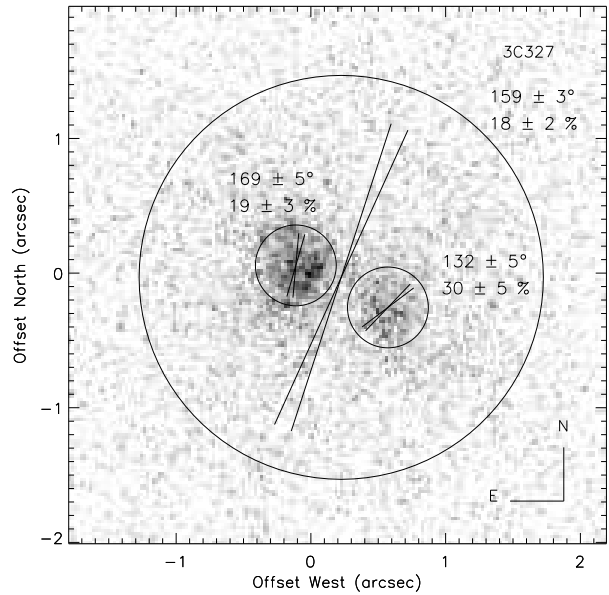}
\figcaption{The same as Fig.\ref{fig-imgpol-3c234}, but for 3C327. The
image is the same as Fig.\ref{fig-aper-c}. \label{fig-imgpol-3c327}}
\end{figure*}

Table \ref{tab-comp} compares our spectropolarimetric results with the
previous observations. Slight discrepancies in position angles (PAs)
are possibly explained through the differences in observing aperture
sizes.  We show the polarization extracted at different positions and
apertures from the HST UV imaging polarimetry data \citep{Hu99} taken
with F320W filter ($\lambda_{\rm obs} \sim 3100$\AA) in Figures
\ref{fig-imgpol-3c234} and \ref{fig-imgpol-3c327}.  In 3C234, the
nucleus has an extension to the south east on $\sim 0.''5$
scale. (Also the nucleus seems to be slightly elongated to the north
east on $\sim 0.''1$ scale. The PSF through the F320W filter is known
to be elongated, but this elongation direction is different from the
one seen in the nucleus.) This extension seems to be polarized
differently from the nucleus, though the values for this component in
Figure \ref{fig-imgpol-3c234} are only suggestive since this FOC image
is pre-COSTAR data.  The optical imaging polarimetry data of
\citet{Co99} show similar PA distribution (see their Fig.5f).  Our FOS
observing aperture has possibly excluded some portion of the
south-east component. This would explain at least partly the slight
difference in PA between our FOS data and larger aperture data.  For
3C327, the nucleus is double, or the image suggests the existence of
two scattering cones. These two features have slightly different
polarization. Our FOS aperture was centered on the brighter one, and
the polarization extracted only for this feature is consistent with
our FOS result. As described in \citet{Hu99}, the exposure time for
one of the polarizer images of 3C327 was uncertain, but the
uncertainty is probably less than 5 seconds (R. Jedrzejewski, private
communication; we adopted the same exposure time as in Hurt et
al. 1999). The resulting uncertainty of the measured polarization in
Figure \ref{fig-imgpol-3c327} is less than the quoted statistical
error. Conversely, this is supported by the agreement with our FOS
result.

\begin{figure*}
\plotone{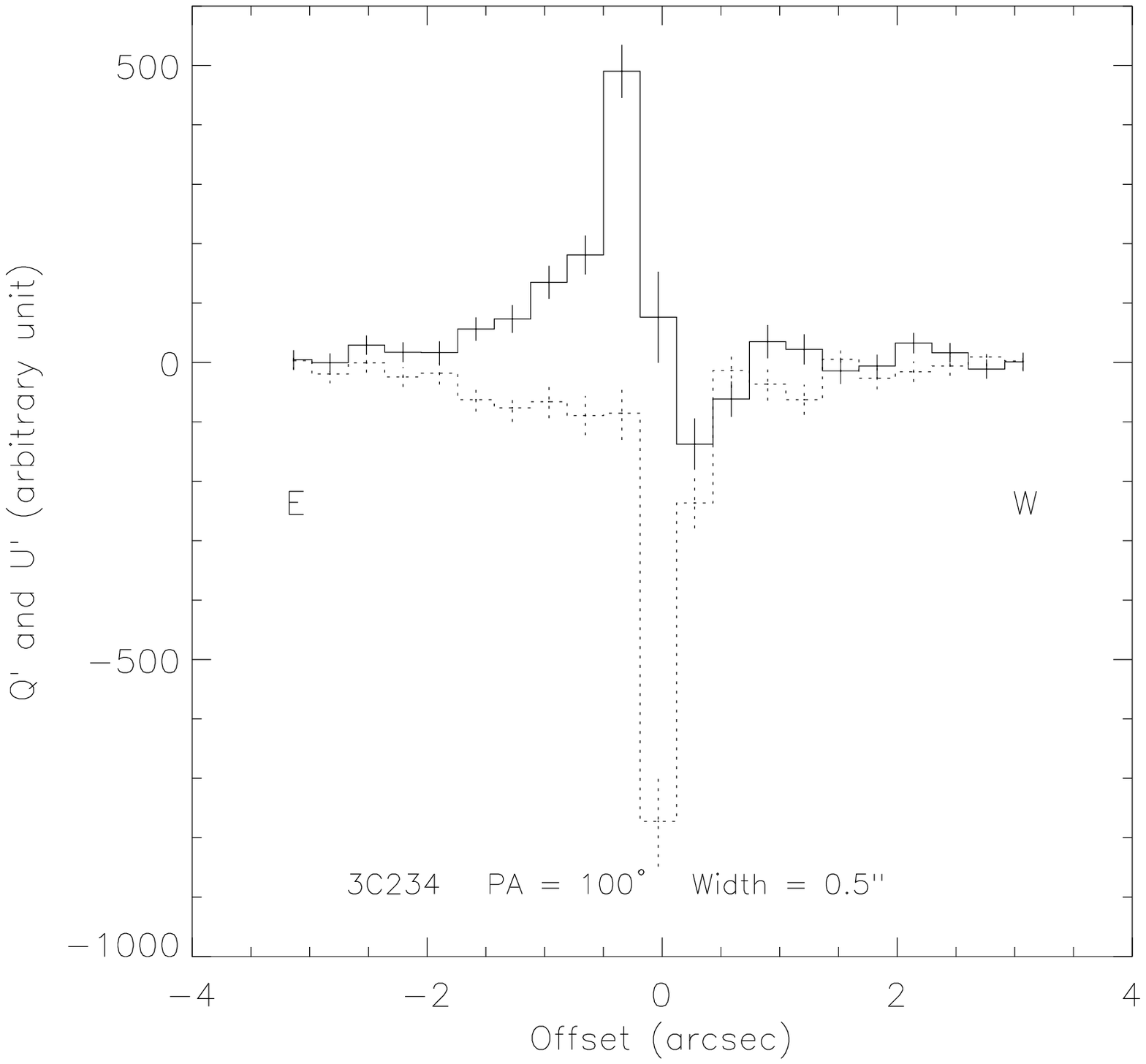} 
\figcaption{Polarized flux distribution in 3C234 along the direction
of PA = 100\degr. The extraction width is $0.''5$. See
Fig.\ref{fig-aper-a} for the extraction window. The solid and dotted
lines indicate Stokes $Q'$ and $U'$, respectively, with its reference
axis perpendicular to PA = 100\degr. Note that the size of the
polarized flux distribution was influenced by the aberrated central
bright point-like source (pre-COSTAR data). \label{fig-pf-a}}
\end{figure*}

From the resolved polarized flux distribution in the imaging
polarimetry data, we can derive the physical extent of the scattering
region.  We have extracted the image with a certain width along the
elongation direction for each object. The Stokes parameters $Q'$ and
$U'$ are calculated with the polarization reference axis perpendicular
to this elongation.  The extracted regions (direction and width) are
indicated in Figures \ref{fig-aper-a}$\sim$\ref{fig-aper-c}, and we
show the $Q'$ and $U'$ distributions in Figures
\ref{fig-pf-a}$\sim$\ref{fig-pf-c}. The figures indicate that the
polarized fluxes from these radio galaxies are extended from a
sub-arcsecond to arcsecond scale, which corresponds to a few kpc to
10kpc (though the size for 3C234 would be lower, since the
imaging polarimetry data are pre-COSTAR). This is one or more orders
of magnitude larger than the spatial scale of the scattering region
seen in Seyfert galaxies. The sizes are summarized in Table
\ref{tab-scatsize}.

\begin{figure*}
\plotone{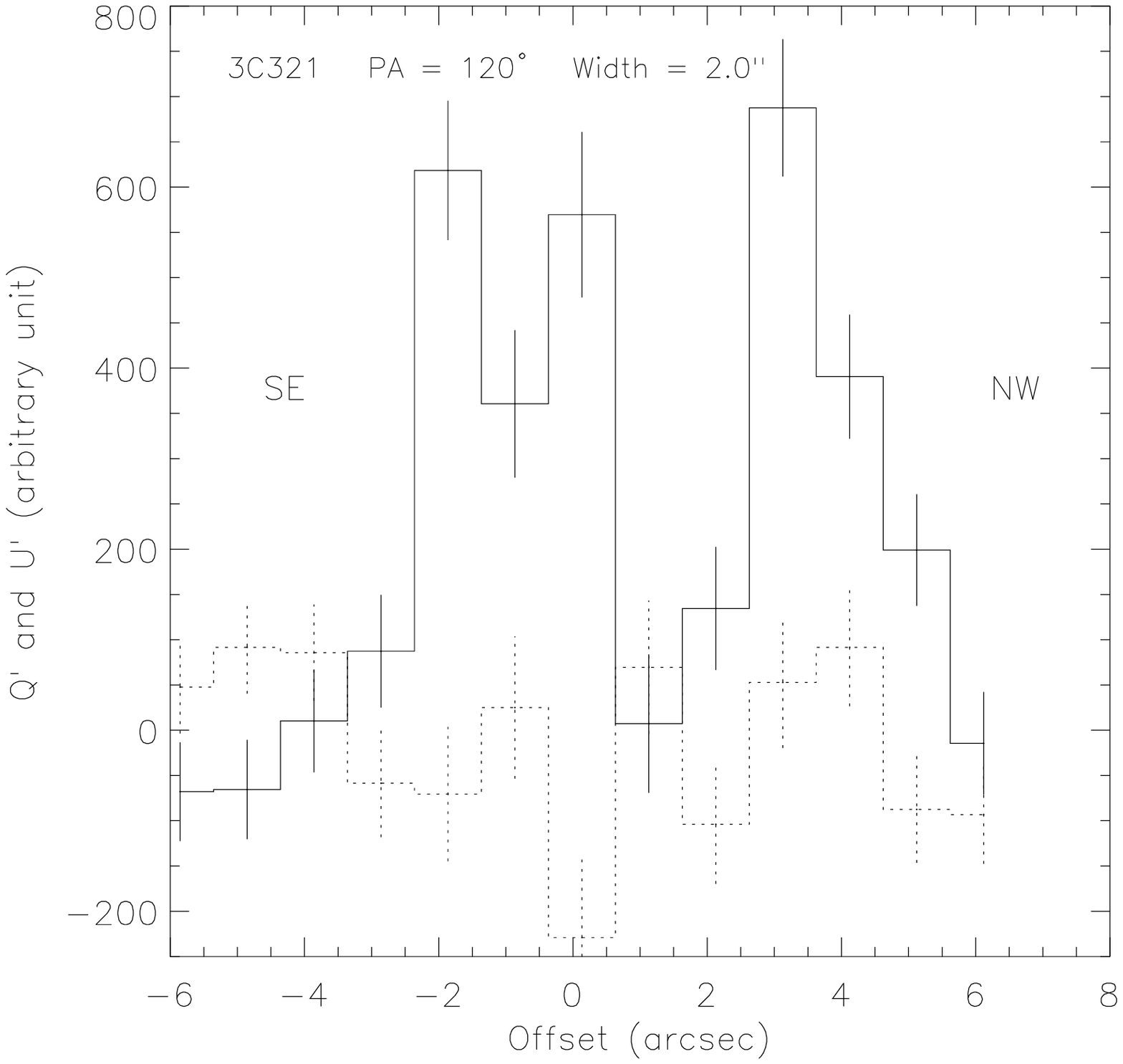} 
\figcaption{ The same as Fig.\ref{fig-pf-a}, but for 3C321 along the
direction of PA = 120\degr and width = $2.''0$. See
Fig.\ref{fig-aper-b} for the extraction window. \label{fig-pf-b}}
\end{figure*}

\begin{figure*}
\plotone{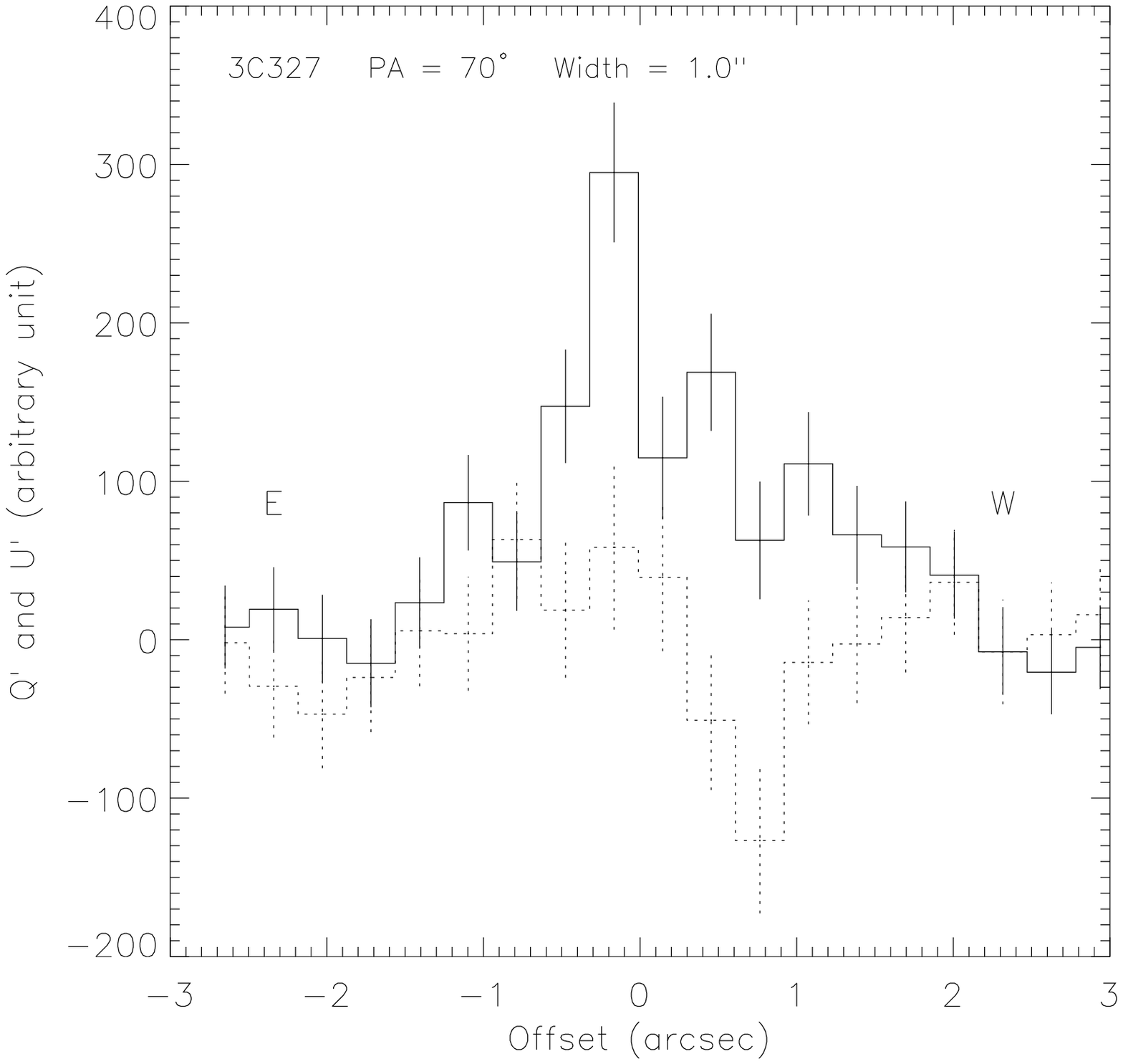} 
\figcaption{The same as Fig.\ref{fig-pf-c}, but for 3C327 along the
direction of PA = 70\degr and width = $1.''0$. See
Fig.\ref{fig-aper-c} for the extraction window. \label{fig-pf-c}}
\end{figure*}

\section{Discussion}\label{sec-disc}

In this section, we discuss the observed polarized flux spectra
comparing these with the observed quasar continuum.  We refer to the
typical quasar continuum shape as $\alpha \simeq -0.3$ where $F_{\nu}
\propto \nu^{\alpha}$ at $\lambda_{\rm rest} = 1 \sim 0.15$\micron\
\citep{Ne87,Fr91,Fr96}.  Some papers find steeper typical slopes, in
part because of the apparent luminosity dependence \citep{MW89}, and
sometimes because the so-called 3000\AA\ bump is not removed.  Also of
course since the underlying continuum is sometimes convex, the rest
wavelength intervals are important.  We note that, however, even if
the intrinsic optical and UV spectral shape ($\gtrsim2000$\AA) is
slightly redder, such as the composite spectra of $\alpha \sim -0.7$
\citep{CV90}, this essentially does not affect our conclusions below.
The spectral shape could be even redder at shorter wavelength
\citep{Zh97}, but this basically is not relevant here since we mainly
discuss the polarized flux of wavelength longer than $\sim 2000$\AA,
except for the discussion of high-redshift radio galaxies (\S
\ref{sec-disc-other}).

\subsection{The feasibility of electron scattering}
\label{sec-disc-elec}

Our UV spectropolarimetry shows that the continuum polarization is
roughly independent of wavelength, and as we will see in
\S\ref{sec-disc-ind}, the polarized flux shape from the UV to optical
is similar to or redder than the typical quasar continuum.  These
might suggest that the dominant scattering is being caused by
electrons (plus reddening), rather than by dust grains. Its
feasibility, however, is quite low.  As we have seen in the previous
section, the scattering regions in these radio galaxies are rather
large.  The previous work on the distant high-redshift NLRGs has shown
that the scattering regions are even larger, $\gtrsim 10$ kpc scale
(3C256 : Dey et al. 1996; 3C265 : di Serego Alighieri et al. 1996,
Tran et al.  1998; 3C324 : Cimatti et al. 1996). This fact favors dust
grains as the dominant scatterers. First, the scattering cross section
per hydrogen atom, $\sigma$, for dust grains in the optical or UV is
two or three orders of magnitude larger than Thomson scattering cross
section in the interstellar medium of our Galaxy (see below for more
specific values), and may not be too different in the scattering
region in these radio galaxies. Dust should survive at these large
scale regions, and large dust masses are observed to exist in many
high-redshift radio galaxies and quasars
\citep{Du94,CK94,Ci98a}. Secondly, even if the amount of dust is a few
orders of magnitude less than that in our Galaxy and the electrons are
to take the role as the dominant scatterers, the large sizes of the
scattering region observed imply enormous masses for the scattering
gas \citep{di94,Ci96,De96}. This argument is summarized as follows.

The scattered luminosity $L_{\nu}^s$ is described by, using the
nuclear luminosity $L_{\nu}$, 
\begin{equation}
L_{\nu}^s = \frac{L_{\nu}}{4\pi} \sigma \int \frac{n}{r^2} dV.
\end{equation}
where $n$ is the hydrogen density, $r$ is the radial distance from the
nucleus, $\sigma$ is again the scattering cross section per hydrogen
atom, and $V$ is the volume. The integral is taken over the whole
scattering region, assuming an optically thin case.  Let us define a
geometrical factor $g$ as
\begin{equation}
g = \frac{\int n dV}{R^2 \int \frac{n}{r^2} dV}
\end{equation}
where $R$ is a characteristic size for the scattering region. This
factor depends on the geometry of the scattering region but is of
order unity. Using this factor, we can relate the total number of
hydrogen $\int n dV$ to the scattering fraction
$L_{\nu}^s/L_{\nu}$ as
\begin{eqnarray}
\frac{L_{\nu}^s}{L_{\nu}} &\equiv& \tau \Omega /4\pi \nonumber\\
&=& g^{-1} \frac{\sigma}{4\pi R^2} \int n dV
\label{eq-reffrac}
\end{eqnarray}
Here, we rewrote the scattering fraction as $\tau \Omega /4\pi$ (which
means that it corresponds to the average optical thickness of the
scattering medium, $\tau$, multiplied by the covering factor of the
medium with respect to the central illuminating source, $\Omega/4\pi$,
but we are not assuming any particular geometry of the scattering
region here). Thus the mass for the scattering gas $M_{\rm scat}$
necessary to achieve a certain scattering fraction is written as
\begin{equation}
M_{\rm scat} = 3.8 \cdot 10^{10} 
\left(\frac{\tau \Omega /4\pi}{0.01} \right) g
\left(\frac{R}{5 {\rm kpc}}\right)^2 
\left(\frac{\sigma}{\sigma_T}\right)^{-1} 
M_{\sun},\label{eq-mscat}
\end{equation}
where $\sigma_T$ is the Thomson scattering cross section. Note that
the gas mass surrounding the nucleus could be larger than $M_{\rm
scat}$ by a factor of several or more, since we are considering only
the gas illuminated by the nucleus and there should be more
unilluminated gas (e.g. outside of the ionization cone).

We have adopted the value of 0.01 as a typical value for the
scattering fraction $\tau \Omega /4\pi$. For our three objects, we can
compare their UV luminosity with that of the quasars with the same
low-redshift range and with the same radio lobe power. Those differ
from our objects mainly in orientation alone according to the unified
model. The UV polarized flux luminosity of our three objects is Log
$PL_{\nu}$ (erg sec$^{-1}$ Hz$^{-1}$) = $27.4\sim28.2$ at
$\lambda_{\rm rest} \sim 2500$\AA\ and their radio power at 178MHz is
Log $P_{178}$ (erg sec$^{-1}$ Hz$^{-1}$ sr$^{-1}$) = $32.6 \sim 33.7$,
while the 3C quasars with $z \lesssim 0.3$ and with the same radio
power range have Log $L_{\nu}$ = $29.0 \sim 30.9$ at $\lambda_{\rm
rest} = 2500$\AA\ (we have taken absolute magnitudes from Veron-Cetty
\& Veron 2000 and converted them with $F_{\nu} \propto
\nu^{-0.3}$). If we think that the intrinsic polarization is less than
$\sim$ 50\%, the scattered light would be larger than the polarized
flux at least by a factor of $\sim 2$. Therefore the scattering
fraction 0.01 is a representative approximation. Consideration of
reddening would make the scattered luminosity of these radio galaxies
brighter, so would make $M_{\rm scat}$ even larger.  Thus, if the
scatterers are electrons $(\sigma = \sigma_T)$, a large mass is
implied.  On the other hand, only a relatively small mass is necessary
if the dominant scatterers are dust grains: the scattering cross
section $\sigma$ for dust grain per H atom is $\sigma \sim 300
\sigma_T$ at $V$ and $\sigma = 600 \sim 900 \sigma_T$ at $\lambda =
3000 \sim 2000$\AA\ in our Galaxy, assuming the column density of H
atoms per $E(\bv)$ = $7.5 \cdot 10^{21}$ cm$^{-2}$ \citep{JS74},
$A_V/E(\bv) = 3.1$, and albedo of 0.5.

The recombination-line luminosity provides another argument, if we
consider the line luminosity from the scattering region.  Let us
consider a simple case of a scattering region with density $n$, volume
$V$, and volume filling factor $f$, so that we have the amount of gas
$\int n dV \simeq n V f$ and the emission measure $\int n^2 dV \simeq
n^2 V f$. With the scattering region size $R$ and the amount of the
gas $\int n dV$ from the above estimation [equation (\ref{eq-mscat})],
we can estimate the \ha\ luminosity $L_{H\alpha}$ to see if we predict
the right amount with a reasonable value of $f$;
\begin{eqnarray}
L_{H\alpha} &\simeq& 4\pi j_{H\alpha}
\frac{ \left(\int n dV \right) ^2 }{ V \cdot f } 
\nonumber\\
&=&  4.7 \cdot 10^{43} \cdot f^{-1} \cdot
\left( \frac{\tau \Omega /4\pi}{0.01} \right)^{2} g^2 \nonumber\\
&& \mbox{ } \left( \frac{R}{5{\rm kpc}} \right)
\left(\frac{\sigma}{\sigma_T}\right)^{-2}
\enspace {\rm erg}~{\rm sec}^{-1}. \label{eq-ha}
\end{eqnarray}
We simply adopted the volume of $\frac{4}{3} \pi R^3$, and the
H$\alpha$ emissivity $j_{H\alpha}$ for $T=10^4$K.  The density is
estimated as
 \begin{eqnarray}
 n &\simeq& \frac{\int n dV}{V f} \nonumber\\
 &=& 2.9 \left( \frac{\tau \Omega /4\pi}{0.01} \right) g
 \left( \frac{R}{5 {\rm kpc}} \right)^{-1} \nonumber\\
 && \mbox{ } \left( \frac{\sigma}{\sigma_T} \right)^{-1} f^{-1}
 \enspace {\rm cm}^{-3}.
 \label{eq-n-f}
 \end{eqnarray}
For our three radio galaxies, H$\alpha$ luminosity is of the order of
$10^{42} \sim 10^{43}$ erg sec$^{-1}$ (table \ref{tab-scatsize}). If
we assume dust scattering, the observed line luminosity can be easily
accomodated with a small value of $f$ ($10^{-4} \sim 10^{-5}$). The density
will be of order $10^2$ cm$^{-3}$ in this case. On the other hand,
when assuming electron scattering, to avoid the overprediction of
H$\alpha$ luminosity, the volume filling factor should be close to
unity, or not be too far from unity, which is quite
surprising.\footnotemark[2] The density will be of order 1 cm$^{-3}$
in this case.

This situation can be summarized as follows.  In the electron case,
for which a large amount of gas is implied, the gas has to be spread
over the entire volume to have smaller density for suppressing the
line emission. In the dust case, for which a smaller amount of gas is
needed, the gas should be squeezed into a smaller effective volume to
have higher density to get the same line emission.

Higher $T$ of $\sim 10^7$ K, as will be discussed in
\S\ref{sec-disc-ind} (but not too high to avoid smearing out the
scattered broad line completely), could reduce $L_{H\alpha}$ by about
a few orders of magnitude.  One possibility for electron scattering is
a hot diffuse volume-filling gas without much line emission, possibly
surrounding the narrow-line region, though this still suffers from the
large mass problem.

\footnotetext[2]{It is conceivable that gas with $n_e \sim 3$
cm$^{-3}$ (see eq.[\ref{eq-n-f}]) and $T_e \sim 10^{4\sim6}$ K does
fill much of the volume, in pressure equilibrium with the hot X-ray
emitting gas expected for an elliptical host. This arrangement is
unstable and inplausible, but is perhaps suggested by the soft X-ray
absorption seen in cooling flow spectra.}

\subsection{Optically thin dust scattering}
\label{sec-disc-thin}

The mostly constant polarization and the polarized flux spectrum
no bluer than typical quasars would lead us to consider electron
scattering, but the feasibility arguments above would suggest dust
grains to be the dominant scatterers. Below we attempt to summarize
the property of dust scattering, for the interpretation of the
polarization data of radio galaxies.

The most robust way to investigate the wavelength dependence of the
scattering process is to look first at the polarized flux spectrum
with a wide wavelength range, then look at the total flux spectrum
considering the possible diluting components.  These diluting
components, such as stellar radiation and nebular continuum, usually
have no influence on the polarized flux spectrum.  The polarized flux
from the scattering process is the product of the incident spectrum
and the polarizing efficiency $\epsilon$, where $\epsilon$ is the
scattering efficiency $\kappa$ times the intrinsic polarization degree
$P_0$ which is the degree of polarization of the scattered light from
the scattering process. In optically thin dust scattering, $\kappa$ is
simply proportional to the scattering cross section $\sigma$ which has
a rather strong wavelength dependence, and $P_0$ has a moderate
wavelength dependence.  We have to compare the product of these two
with the ratio of the observed polarized flux spectrum to the typical
quasar continuum shape, $\alpha \simeq -0.3$ where $F_{\nu} \propto
\nu^{\alpha}$. For the optically thick case, $\kappa$ will be
significantly different, which will be discussed in the next section.
We focus on optically thin cases in this section.

\begin{figure*}
\plotone{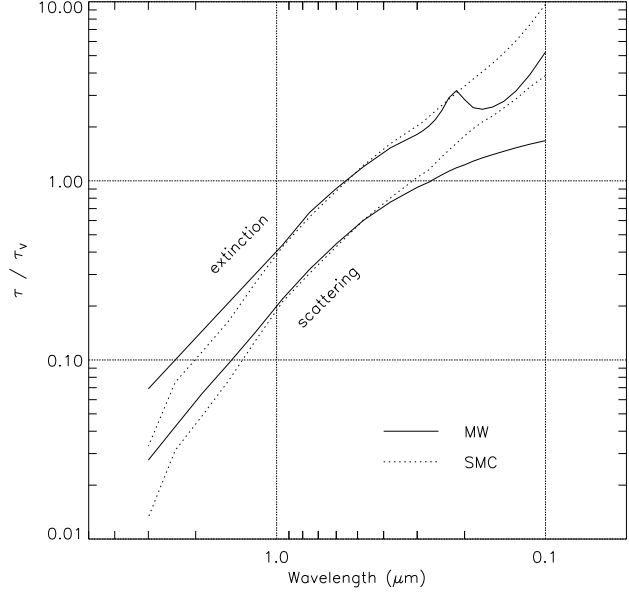}
\figcaption{The wavelength dependence of extinction (upper curve) and
scattering (lower curve) optical depth for Galaxy (solid) and SMC
(dotted). The data values are taken from
\citet{WG00}.\label{fig-crosssec}}
\end{figure*}

The scattering cross section of the extreme case is the Rayleigh
scattering limit where $\sigma \propto \nu^{+4}$.  This is for the
case of very small particles. For Galactic dust, the scattering cross
section has been investigated through the observations of reflection
nebulae (Witt \& Gordon 2000 and references therein). It generally
increases with decreasing wavelength, but its wavelength dependency
changes with wavelength range, as shown in Figure
\ref{fig-crosssec}. The scattering cross section is proportional to
$\sim \nu^{+2}$ in the near IR ($\sim 2\ \micron - 0.9\ \micron$) and
its spectral index gets flatter with shorter wavelength. In the
optical ($\sim 0.9\ \micron - 0.3\ \micron$) it is roughly
proportional to $\nu^{+1}$. In the UV, the {\it extinction} cross
section has a bump at $\sim$ 2200\AA, but this feature seems to be
purely in absorption, not in scattering \citep{Cal95}. This 2200\AA\
feature is weak in the LMC (Misselt, Clayton, \& Gordon 1999 and
references therein), and it is absent in the star-forming bar of the
SMC (Gordon \& Clayton 1998 and references therein) and in starburst
galaxies \citep{CKS94,GCW97}.  In this UV range ($\sim 0.3\ \micron -
0.1\ \micron$), while the extinction cross section for Galactic dust
generally goes up rapidly at wavelengths shorter than the bump
(although it depends on the sight line: Calzetti, Clayton, \& Mathis
1989), the wavelength dependence of the scattering cross section seems
to become somewhat flatter with $\sim \nu^{+0.5}$, though the
empirical work on the albedo at $\lambda \lesssim 2000$\AA\ varies for
different objects and thus it is still poorly understood
\citep{WG00}. For the SMC, the extinction rapidly goes up roughly with
$\nu^{+1.5}$, but no empirical work on the scattering cross section
exists.

Several authors have done theoretical calculations of dust properties.
In an early paper, \citet{MRN77} inferred the composition and size
distribution of the dust grains that reproduce the Galactic extinction
curve. \citet{W79} calculated the scattering and polarization
properties for this population of dust. Recently \citet{ZL00} have
calculated the polarization properties using updated dielectric
functions. These latter two calculations suggest that the polarization
of the scattered light in a single dust scattering process, $P_0$, is
moderately wavelength-dependent. In \citet{ZL00}, the maximum
polarization, which is achieved when the scattering angle is around
90\degr, is approximately flat at about 30 $\sim$ 35\% (versus 100\%
for electron scattering) in the UV/optical region of $\sim$ 1700\AA\
$-$ 7000\AA. To longer wavelength, it goes up to $\sim60$\% at $\sim
1\ \micron$. To shorter wavelength in the far UV, it also rather
rapidly increases reaching $\sim60$\% at 1000\AA. The region around
2200\AA\ would be highly uncertain because these theoretical
calculations produce the 2200\AA\ enhancement in the scattering cross
section, as opposed to the observational result by \citet{Cal95},
though this is the only empirical result at present.  At smaller or
larger scattering angle, the polarization degree becomes smaller, but
the overall wavelength dependence is similar.  We obtain the
polarizing efficiency $\epsilon$ by multiplying the scattering cross
section and this polarization degree $P_0$.  For the region of the
most interest in this paper, $0.2 - 0.7$\micron, $P_0$ seems to be
approximately constant, so the shapes of the scattering efficiency
$\kappa$ and polarizing efficiency $\epsilon$ are almost the same, and
thus $\epsilon$ is proportional to roughly $\nu^{+1}$ for the Galactic
dust, if there is no 2200\AA\ bump in the scattering cross section as
observed by \citet{Cal95}. For the SMC dust, $\epsilon$ would be
roughly proportional to $\nu^{+1.5}$. For the shorter wavelenth
(far-UV), $\epsilon$ would be slightly bluer, and for the longer
wavelength (near-IR), it would be somewhat redder.

Some observational results actually show this blue polarizing
efficiency or blue scattering efficiency. In the prototype Seyfert 2
galaxy NGC 1068, an off-nuclear knot at about $5''$ NE from the
nucleus shows a blue polarized flux spectrum $\alpha \simeq +1$ where
$F_{\nu} \propto \nu^{\alpha}$ ($\beta \simeq -3$ where $F_{\lambda}
\propto \lambda^{\beta}$; Fig.5 in Miller, Goodrich, \& Mathews 1991)
in the optical region ($\lambda\lambda 3600-6400$\AA). We can compare
this spectrum with the polarized flux from the nuclear vicinity to
obtain the polarizing efficiency, since the scatterers are considered
to be electrons there. Its spectral index is $\alpha \simeq -0.5$
($\beta \simeq -1.5$; Fig.6 in Miller et al. 1991) or $\alpha = 0.0$
($\beta=-2.0$; with a wider wavelength coverage and a narrower slit
width; Tran 1995), so we obtain the polarizing efficiency in this case
to be $\epsilon \propto \nu^{+1} \sim \nu^{+1.5}$. This is consistent
with the general optically-thin dust scattering property.  In the
radio galaxy PKS2152-69 \citep{di88}, the extranuclear cloud shows
very blue scattered light, $\alpha = +3 \pm 1$, which could indicate
that the scattering efficiency (not the polarizing efficiency in this
case) $\kappa$ is roughly proportional to $\nu^{+3 \pm 1}$, assuming a
typical quasar spectral index for the nuclear continuum. This could be
close to the case of Rayleigh scattering, but with a rather large
uncertainty. None of other radio galaxies is known to show this type
of behavior.

\subsection{Opaque clumpy dust clouds}
\label{sec-disc-thickdust}

What if the dust clouds are optically thick ?  Generally, if we see a
scattering wall or slab of dusty gas which is optically thick at all
wavelength regions, the wavelength dependence of the scattering
efficiency would essentially correspond to the albedo (not the
scattering cross section) of the scattering medium. For Galactic dust,
the albedo is observed to be mostly flat in the UV-IR wavelength
regions. In this sense, its scattering efficiency curve would be
rather similar to electron scattering.  But in reality, we will
observe the scattering region through various viewing angles, and this
would cause a wavelength dependence of the scattering efficiency
especially when the clouds are only moderately optically thick.  Also,
in the case of radio galaxies, the scattering region would not be
uniformly optically thick as a whole in terms of the unification of
radio galaxies and quasars, since otherwise we would not observe any
quasars.  The optically thick regions should be localized, i.e., the
scattering region as a whole should be clumpy as actually observed in
3C321, and it is each of these clumps which could be optically
thick. Then, the general property of this scattering region can be
approximately derived from that of a single spherical scattering blob.

\begin{figure*}
\plotone{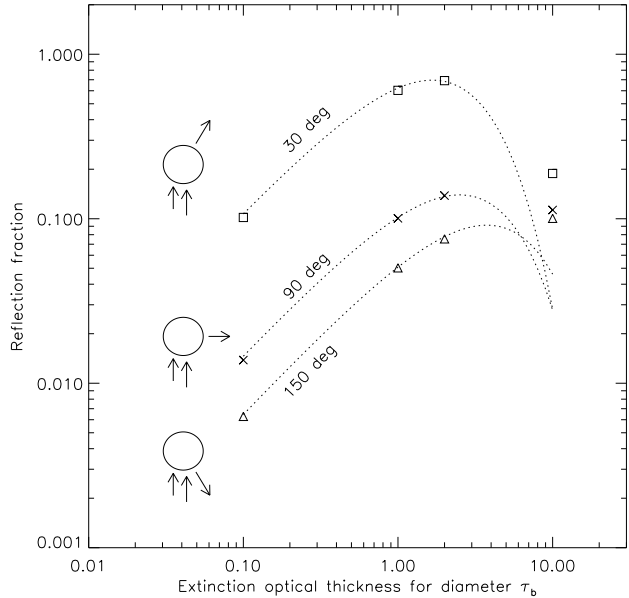} 
\figcaption{Scattered flux from a spherical dusty blob as a function
of extinction optical thickness for diameter $\tau_b$ and for
different viewing angles (which correspond to the scattering angle at
the blob), taken from \citet{CW95}. It is the scattered flux per unit
solid angle with the incident flux being unity, multiplied by $4\pi$.
Therefore it corresponds to the scattering fraction for a blob
integrated over the whole solid angle assuming isotropic scattering.
The squares, crosses, and triangles are for the viewing angle 30\degr,
90\degr, and 150\degr, respectively. The dotted lines illustrate the
fits by the form of $\tau_{\rm scat} \exp (-\tau_{\rm ext}^{\rm eff})$
for small $\tau_b$, which fail at larger $\tau_b$.
\label{fig-dustblob}}
\end{figure*}

The calculation of the scattering properties of such a blob has been
implemented by \citet{CW95} using the Monte Carlo method.  Figure
\ref{fig-dustblob} reproduces their result for the dust scattering
with albedo = 0.5 and scattering asymmetry parameter $g$ = 0.56
(forward-throwing), which are for a representative case for UV/optical
dust scattering \citep{WG00}. The symbols in Figure \ref{fig-dustblob}
are the scattered flux per unit solid angle for a certain viewing
angle $\psi$ (which is the scattering angle at the blob) with unit
incident flux, multiplied by $4\pi$. It corresponds to the scattering
fraction for a blob, integrated assuming isotropic scattering. Since
the single dust scattering process is forward-throwing, the scattering
dusty blob is also forward-throwing, and this is true even for the
case of rather large optical thickness. (We multiplied the
scattered flux by $4\pi$ to show an {\it apparent} scattering
fraction, which we would infer from observations because viewing angle
is usually unknown.)

When the viewing angle $\psi$ is small (forward scattering case), we
will see the strong forward-scattered light through a significant
amount of extinction. When viewed edge-on ($\psi \sim 90\degr$), this
extinction effect becomes smaller. When $\psi$ is large (backward
scattering case) we will see relatively weak back-scattered light with
a smaller amount of extinction, and the scattered light will be
saturated with large optical thickness, just like seeing a thick wall.
In all these cases, the scattered flux first increases proportionally
to the increase of the scattering optical thickness of the blob
(denoted here as $\tau_{\rm scat} \equiv a \tau_b$, where $a$ is
albedo and $\tau_b$ is extinction optical thickness for the diameter
of the blob). Then it starts to have some effective amount of
extinction ($\tau_{\rm ext}^{\rm eff}$), which increases with the blob
optical thickness ($\tau_{\rm ext}^{\rm eff} \propto \tau_b$).  For
relatively small blob optical thickness, the scattered flux can be
approximated by $\tau_{\rm scat} \exp (-\tau_{\rm ext}^{\rm eff})$
(similar expressions have been used by other authors, e.g. Cohen et
al. 1999).  More specifically, the data points, say $f$, in Figure
\ref{fig-dustblob} for relatively small $\tau_b$ can be approximated
by
\begin{equation}
f = a (2\tau_b /3) \cdot 4 \pi \Phi(\psi,g) \cdot 
     e^{-\tau_{\rm ext}^{\rm eff}},
\end{equation}
where
\begin{equation}
\tau_{\rm ext}^{\rm eff} = \xi (2\tau_b /3).
\label{eq-ext-eff}
\end{equation}
The phase function $\Phi$ is given in \citet{HG41}. The factor 2/3
added to $\tau_b$ represents the effective thickness of the sphere of
diameter $\tau_b$.  The factor $\xi$ corresponds to the effective
fraction of extinction of the scattered light. The dotted lines in
Figure \ref{fig-dustblob} illustrate that we can fit the calculation
results with only one parameter $\xi$ for relatively small $\tau_b$
(up to $\sim$ 2.0). This effective extinction fraction $\xi$ is close
to unity but decreases with increasing viewing angle (the adopted
values for the fit in Fig.\ref{fig-dustblob} are 0.9, 0.6, 0.4 for
$\psi = 30\degr, 90\degr, 150\degr$), since we will see the light with
a shorter path inside the blob for larger viewing angle.

In terms of the observations, this $\tau_b$ dependence of the
scattered light corresponds to the wavelength dependence of the
scattering efficiency since the optical thickness of the blob is
proportional to $\sim \nu^{+1}$.  It initially follows the form of
$\tau_b e^{-\xi(2\tau_b/3)}$. However, for larger optical thickness,
which corresponds to shorter wavelength observation, the scattered
flux will not decrease exponentially, as illustrated in Figure
\ref{fig-dustblob}. The fit described above fails at larger optical
thickness. This is because for larger $\tau_b$ the effective
extinction optical thickness $\tau_{\rm ext}^{\rm eff}$ does not grow
proportionally to $\tau_b$, but instead grows slower: the parameter
$\xi$ [eq.(\ref{eq-ext-eff})] for a given viewing angle is not
actually constant, but decreases with $\tau_b$ (we only see more
foreground part of the blob where extinction is smaller).  Thus, once
the blob becomes opaque for the observing wavelength, its scattering
efficiency becomes flat or decreases to the shorter wavelength but not
exponentially.  This means that the wavelength dependence of the
optically-thick scattering efficiency is bluer than that of foreground
screen extinction, for which we expect an exponential cut-off at the
shorter wavelength.

We note that in the optically thick case with the Galactic dust, the
scattering efficiency and polarizing efficiency will have some dip
across the 2200\AA, since \citet{Cal95} showed that the albedo has a
dip at the 2200\AA, based on the enhancement in absorption and
non-enhancement in scattering across the 2200\AA\ feature.  The
scattered light (and polarized flux) will decrease around 2200\AA\
because of this absorption enhancement.  In the SMC or starburst
galaxies, the 2200\AA\ feature does not show up in the extinction, so
the albedo would be flat across the 2200\AA\ region. There would be no
feature in the scattered light from such dust grains. The observations
of radio galaxies are still not definitive on this issue. In our
spectrum of 3C234, we do not see any feature. In the spectrum of
3C327, the feature, if it exists at all, is not clear.\footnotemark[3]

\footnotetext[3]{Some evidence for this feature was discussed in
\citet{Ve99}.}

\subsection{Nature of the scattered light in 3C234, 3C321, and 3C327}
\label{sec-disc-ind}

To investigate the nature of the scatttered light for the three radio
galaxies in our program, we combine our UV polarization data with the
previous optical/IR polarimetry data in the literature. The HST UV
imaging polarimetry is available in \citet{Hu99} for all three
objects. The polarized flux spectra from UV to IR are shown in Figures
\ref{fig-comp-a}, \ref{fig-comp-b}, and \ref{fig-comp-c}.

\subsubsection{3C234}

\begin{figure*}
\plotone{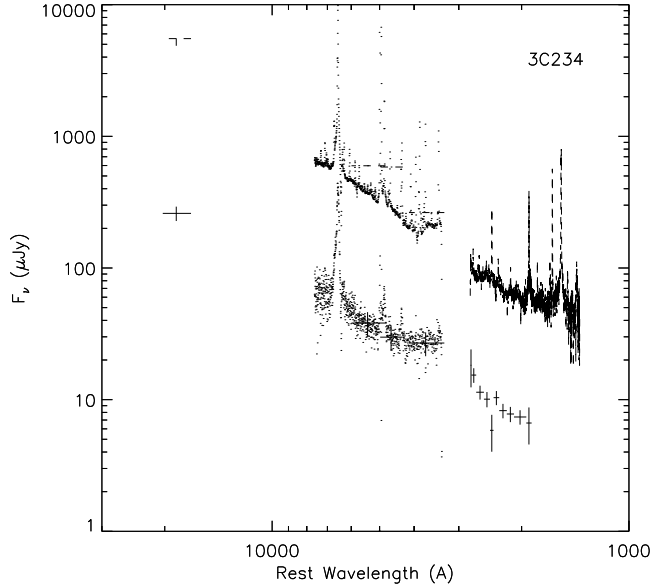}
\figcaption{Total flux and polarized flux observed in the
UV/optical/IR for 3C234. The FOS UV polarized flux spectrum is plotted
in solid ticks with vertical error bars, and above this, the UV total
flux is shown in dashed lines. The optical spectropolarimetry data
\citep{Tr98} are plotted in dots.  The optical/IR broad band data with
large apertures in Table \ref{tab-comp} are shown in solid (polarized
flux) and dashed (total flux) ticks with error bars. The polarized
flux of Tran et al. has been scaled to match the optical
large-aperture polarized flux (so the latter is almost embedded in the
former and hard to see), and the total flux spectrum was also scaled
by the same factor.  All polarized and total fluxes have been
corrected for the small Galacic reddening with $E(\bv) = 0.019$.  
\label{fig-comp-a}}
\end{figure*}

For 3C234, we have plotted ground-based large-aperture polarimetry
data (see Table \ref{tab-comp}) and spectropolarimetry data (Tran et
al. 1995). The latter is a $1''$ slit observation with the seeing
$\lesssim 1''$ and thus the absolute flux is lower, so we have scaled
the data to fit the large aperture data with the same wavelength
regions (of course this is only meaningful if the scattering mechanism
is the same). Although our UV polarized flux with the $0''.86$
aperture could be missing some part of the whole polarized flux, as
indicated by the slight PA rotation (see \S\ref{sec-res}), its flux
level matches that from the large synthetic aperture photometry of the
UV imaging polarimetry data \citep{Hu99}. In Figure \ref{fig-comp-a},
all fluxes have been corrected for the small Galactic reddening
$E(\bv) = 0.019$ (NED; based on Schlegel et al. 1998).

The optical polarized flux has been found to be very red compared to
the typical quasar continuum (Tran et al. 1995) : we measure the
spectral index $\alpha$ ($F_{\nu} \propto \nu^{\alpha}$) between \ha\
and \hb\ to be $\simeq -1.7$ from Tran et al. data (for
$\lambda\lambda5200-5800$\AA\ to avoid any possible influence of broad
\ha\ line).  Also, the Balmer decrement in the polarized flux is large
(7.3; Tran et al. 1995).  These facts, as well as the large \paa/\ha\
ratio for the broad components in the total flux, have been
interpreted to be due to a significant reddening of $A_V = 2.4 \sim
2.7$ (Carleton et al. 1984; Hill et al. 1996; Tran et al. 1995; we
have summarized the observed slope and line ratios in Table
\ref{tab-scatsize}.)

However, a simple screen reddening does not explain our UV polarized
flux observation. The polarized flux spectrum can be fitted with a
power law down to the UV, if we exclude 2600\AA\ $< \lambda_{\rm rest}
< 4000$\AA\ region to allow for the possible so-called 3000\AA\ bump :
$\alpha = -1.7\pm0.2$ for the optical-UV from the large-aperture
optical data and our UV data, and $\alpha = -1.7\pm0.6$ for the UV
only.  A significant amount of dust screen (such as the one to redden
the optical spectrum from $\alpha \simeq -0.3$ to $-1.7$) should
absorb the UV greatly so that the resulting spectrum would have an
exponential decrease toward the UV.  Thus, a dust screen is unlikely
to be the cause of the red continuum of 3C234.

We note that the redness of the optical continuum and Balmer decrement
in the broad-line radio galaxies and radio-loud quasars have been
found to have a correlation, which has been interpreted to be due to
reddening \citep{Baker97}. The observed values for the optical
polarized flux of 3C234 actually fit into this correlation (see Fig.16
in Baker 1997). But our UV data for 3C234 is not compatible with this
reddening interpretation because of the lack of a cutoff in the
UV. Also, note that the reddening inferred in the quasar colors must
show an exponential UV cutoff, if a dust screen is present.

If the polarized flux continuum of 3C234 is without a significant
reddening, this would mean that the broad \ha\ and \hb\ lines in 3C234
have a different reddening. The possibility of Broad lines having
different reddening from continuum has been discussed by some authors
(e.g. Baker 1997). Alternatively, the Balmer decrement could be
intrinsically large in 3C234: the optical depth effect will make the
Balmer decrement larger (e.g. Kwan and Krolik 1981). In this case, to
explain the large \paa/\ha\ ratio observed, we might have to consider
a different component in the IR. We will discuss this later below.

As for the scattering mechanism in 3C234, the red, apparently
no-reddening continuum suggests that an optically-thin dust scattering
is not a dominant contributor.  Otherwise, the incident continuum has
to be redder than the observed one, which would be unlikely. On the
other hand, an opaque dust scattering would explain the
observation. If the viewing angle is rather small, it will make the
continuum redder but without an exponential decrease, as we have seen
in the previous section.

As described in \S\ref{sec-res}, however, an extremely broad component
with FWHM of $\sim 20000$ km sec$^{-1}$ is observed in the \ion{C}{4}
line of our UV total flux spectrum (Fig.\ref{fig-line-3c234}), in
addition to an ordinary broad component. We have inspected the \ha\
region in the polarized flux spectrum of Tran et al., and tentatively
identify an extremely broad component of similar width (or even
broader).  These extremely broad lines could be intrinsic to the broad
line region of 3C234, but also could be due to scattering broadening
by hot electrons with $T \sim 10^7$ K. Thus some part of the scattered
light might be from electrons. We note that the gas mass implied for
electron scattering case would be less of a problem for 3C234, because
the observed scattering region is rather compact.

Now we discuss briefly the IR polarized flux.  We have calculated the
K band polarized flux using the polarization data at 2\micron\ with
$5.''8$ aperture \citep{SZ91} and total flux data with $7.''5$
aperture \citep{LLM85}. The polarization data from \citet{Yo98} is
consistent with the former, but the absolute flux seems to be lower by
a factor of $\sim$ 2 due to the smaller aperture ($3.''08 \times 3''$)
as described by them.  We tentatively do not use the H band data from
\citet{Br90}, since the H band total flux ($\sim 1360\mu$Jy) shows a
rather strange dip by a factor of about 2 between the J band and K
band flux (2050 $\mu$Jy and 5500 $\mu$Jy, respectively). Note,
however, that there are emission line contributions in J and K
bands, such as \ion{He}{1} $\lambda$10830 and \paa.

Having these in mind, the observed K band polarized flux forms almost
the same spectral shape as the optical ($\alpha = -1.6\pm0.1$ over the
IR to the optical).  Typical quasar continua actually turn up at
around 1 \micron\ and have a steeper spectral shape in the IR with
$\alpha = -1.4$ (or $-2 \sim -1$; Neugebauer et al. 1987) than in the
optical. Apparently, the IR continuum polarized flux of 3C234 could be
consistent with no or little reddening. However, the large \paa/\ha\
ratio would not be consistent with this. Again, the broad lines would
have different reddening than the continuum, or alternatively, the
large ratio might be suggesting that another polarized component is
coming in in the IR, which could be highly-reddened scattered light
from the inner region, or a dichroic polarization component
\citep{Yo98}. Note that the PA does not change in the IR from the
optical. 

In summary, for the scattering mechanism in 3C234, we infer that
opaque dust scattering with a rather small viewing angle might be
playing a role, though some part of the scattered light could be from
electrons. 

The hybrid case of dust-scattered and electron-scattered light seems
to be realized actually in nature. It could be the situation exactly
in NGC 1068 for a large aperture $6''\times12''$, including the NE
dust knot, observed in the UV \citep{Co93}, where the UV polarized
flux is significantly influenced by the dust scattered light. In the
case of Cygnus A, the polarized H$\alpha$ line is extremely broad
\citep{Og97}, but the \ion{Mg}{2} line is much narrower
\citep{An94}. Therefore, scattering agents could be different in these
two wavelength regions, and electron scattering might be contributing
at least at $H\alpha$, though the Galactic-reddening corrected flux
ratio of these two lines could be consistent with optically-thin dust
scattering \citep{Og97}.

\subsubsection{3C321}

\begin{figure*}
\plotone{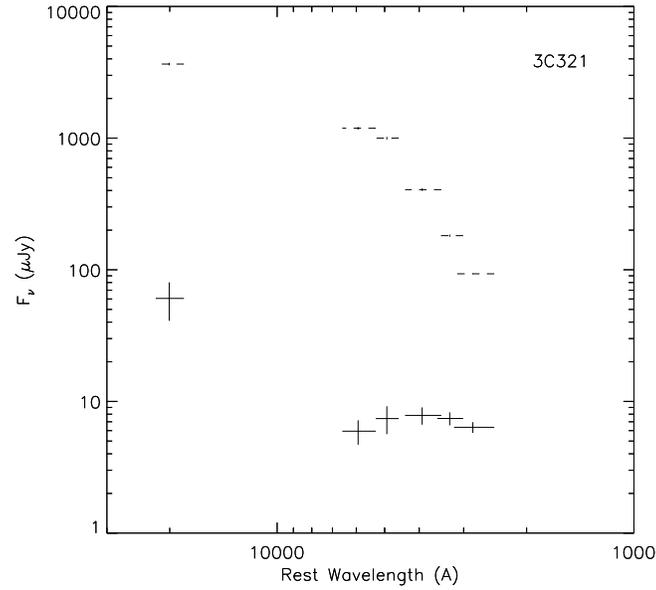} \figcaption{Total flux (dashed ticks) and
polarized flux (solid ticks) observed for 3C321 in the
near-UV/optical/IR wavelength regions are plotted with error bars. All
fluxes have been corrected for the Galactic reddening with $E(\bv) =
0.044$.  The data are in Table \ref{tab-comp}. \label{fig-comp-b}}
\end{figure*}

For 3C321 (Fig.\ref{fig-comp-b}), we have combined the optical/IR
polarization data by \citet{Yo96a} with the FOC UV imaging polarimetry
data. The nucleus of 3C321 consists mainly of two components,
separated by $\sim 3.''3$. The $5''$ aperture polarimetry of
\citet{Yo96a} seems to be centered on the brighter, SE component.
\citet{Hu99}, however, give the integrated polarization for the whole
of both two components, so we have implemented the synthetic $5''$
aperture polarimetry centered on the SE component, the result of which
is shown in Table \ref{tab-comp} and Figure \ref{fig-comp-b}.

The Galactic reddening is $E(\bv) = 0.044$ (NED; based on Schlegel et
al. 1998). The centro-symmetric pattern of the position angle
distribution of the optical polarization does not seem to be disturbed
by the Galactic interstellar polarization \citep{Co99,Dr93}.  The
spectral shape of the polarized flux with the $5''$ aperture corrected
for the Galactic reddening is $\alpha = -0.1 \pm 0.3$ ($\lambda_{\rm
rest} = 2500 \sim 6500$\AA).  The spectropolarimetry by \citet{TDS96}
shows that the optical polarized flux spectrum has $\alpha = -0.3$,
and the polarized flux continuum of \citet{Yo96a} is consistent with
this result.  \citet{Co99} have taken optical spectropolarimetric data
with a slit across these two nuclei ($1'' \times 17''$ extraction
size), and showed that the polarized flux spectrum integrated is of
$\alpha = 0.0$, while the Balmer decrement seems to be not too large.

The scattering region of 3C321 is very extended, so the results of
polarimetry will be different for different size and location of the
apertures and seeing conditions.  Nevertheless, these results suggest
that, overall, the wavelength dependence of the polarizing efficiency
is apparently quite flat or slightly blue in the optical to the
near-UV.

What is the nature of this polarized flux ?  At the central region,
there is evidence for reddening in the morphology of the optical image
taken by HST (see Fig.2 in Hurt et al. 1999) where a dust lane is
clearly seen.  Also, the narrow line components have been shown to be
reddened at the central region : from H$\gamma$/H$\beta$, Tadhunter et
al. (1996) obtained $A_V = 2.9\pm0.6$ with $1.''6$ slit; from
Pa$\alpha$/H$\alpha$/H$\beta$, \citet{Hi96} obtained $A_V = 1.5\pm0.3$
for the central $\sim 2.''5 \times 2.''5$ region.  Therefore, the
polarized flux from this central region could also be reddened.  The
significant part of the polarized flux, however, is coming from the
outer region.  Using the UV imaging polarimetry data of \citet{Hu99},
we have compared the polarized flux from the central $2.''5$ region
with that from the outer region within the $5''$ diameter masking the
central $2.''5$. They are found to be approximately of the same
amount. Therefore, the possible explanation of the flat or slightly
blue polarizing efficiency would be that the reddened polarized flux
from the central $\sim 2''$ region is compensated by the blue
optically-thin dust scatterered light from the outer region.

We also compared the polarized flux from the observed clumps
(including the NW component) with that from the surrounding diffuse
gas, and found that they have a comparable amount, though the
definition of the diffuse region is somewhat arbitrary and also the
data suffer from the abberation (pre-COSTAR). This polarized flux from
the diffuse region would be optically-thin dust scattered light, while
the polarized flux from the clumps might be opaque dust scattered
light. The latter would have roughly neutral scattering efficincy if
the viewing angle is close to edge-on, which would be expected from
the morphology with a dust lane.

However, the possibility of electron-scattering contribution cannot be
ruled out, aside from the huge mass implied if this is the case for
the diffuse extended region. The spectral shape of the polarized flux
is not blue enough to rule it out to be neutrally scattered
light. Also, an inspection of Fig.2 of \citet{Co99} possibly suggests
a faint but extremely broad component at \ha\ [like Cyg A (see the
same Fig.2 in Cohen et al; Ogle et al 1997); or 3C234 (see previous
section)].

The IR polarized flux (see Fig.\ref{fig-comp-b}) seems to have a
different component, though the S/N for the K band data is only about
3. This component could be another highly-reddened scattered component
or could be a dichroic polarization component.  We need more IR
polarimetric observations to clarify the nature of this component.

\subsubsection{3C327}

\begin{figure*}
\plotone{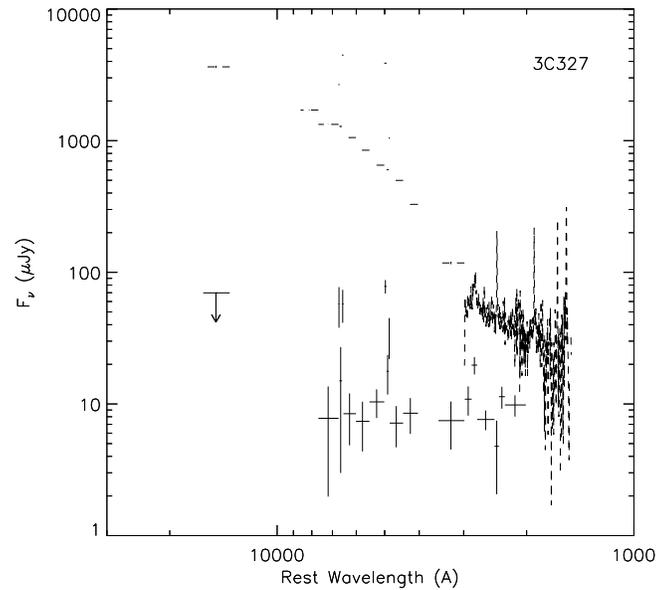} 
\figcaption{Total flux (dashed ticks/lines) and polarized flux (solid
ticks) observed for 3C327 in the UV/optical/IR wavelength regions. The
arrow is a 3$\sigma$ upper limit. The spectra of our FOS observations
have been scaled to match at $\lambda_{\rm obs} = 2700\sim3300$\AA\ to
the FOC imaging polarimetry data.  The unpublished optical
spectropolarimetry data (Dey et al.) are included, which is scaled to
match the 3$\sigma$ upper limit of the polarized flux at B band
(Young et al. 1996b ; see Table \ref{tab-comp}). All fluxes are
corrected for the Galatic reddening with $E(\bv) =
0.089$. \label{fig-comp-c}}
\end{figure*}

The UV and near-UV radiation of 3C327 is highly polarized, while in
the optical the polarization is rather low. In table \ref{tab-comp},
we list our UV data, UV imaging polarimetry data of \citet{Hu99}, and
also the large-aperture optical and IR data of \citet{Yo96b} with
3$\sigma$ upper limit for $P < 2\sigma_{P}$.  In Figure
\ref{fig-comp-c}, we have plotted our UV data scaled by the ratio of
the polarized flux in the large synthetic aperture data of
\citet{Hu99} to our polarized flux with $0.''86$ aperture at
$\lambda_{\rm obs} = 2700\sim3300$\AA, assuming that the two regions
seen in the UV image have the same polarized flux spectral shape. This
scaling ratio was found to be 2.27.  The U band and H band data from
\citet{Yo96b} are also shown, with the latter as an upper limit. To
illustrate the optical polarized flux in Figure \ref{fig-comp-c}, we
have included optical spectropolarimetry data obtained at the Keck
telescope during a cloudy night in June 1996 [$1''$ slit, 10 minutes
exposure per position angle; Dey, Antonucci, \& Cimatti, unpublished.
Note that [\ion{O}{3}]$\lambda$4959+5007 seems to be detected in the
polarized flux, which was first suggested by \citet{di97}]. We scaled
the data to have the polarized flux at the blue end ($\sim4700$\AA)
match with the 3$\sigma$ upper limit at the B band. With this scaling,
the total flux is only slightly smaller ($10\sim20$\% level) than the
large-aperture data.

If we correct for the Galactic reddening, taking $E(\bv) = 0.089$
(NED; based on Schlegel et al. 1998) and $R_V = 3.1$, then we obtain
$\alpha = +0.5\pm0.8$ for the polarized continuum (using continuum
bins only) from the UV to near-UV ($\lambda_{\rm rest} = 2000 \sim
3500$\AA; FOS + U band data).  The optical polarized flux spectrum,
corrected for the Galactic reddening assuming negligible contribution
from interstellar polarization, is consistent with this result at the
UV, though not too constraining : the spectral index $\alpha$ is
$-0.01\pm0.94$. If we combine the UV and optical polarized flux, we
obtain $\alpha =-0.01\pm0.21$ ($\lambda_{\rm rest} = 2000 \sim
7500$\AA), though there is uncertainty from the scaling and using the
upper limit at B band (the true shape could be bluer). There could be
an influence from the 3000\AA\ bump, and in that case, the intrinsic
shape would be slightly redder.

The narrow lines have been found to be reddened.  From
Pa$\alpha$/H$\alpha$/H$\beta$ ratios, \citet{Hi96} obtained $A_V=1.2
\pm 0.3$ (see Table \ref{tab-scatsize} for the observed ratios). The
narrow emission line region is observed to be extended (total extent
of $\sim 4''$; Baum et al. 1988), though not well resolved.  On the
other hand, the polarized flux does not seem to show a large decrease
in the shorter wavelength. Therefore, the scattering region may not be
cospatial with the narrow-line region, or again, the optically-thick
dust clouds with an edge-on view (as inferred from the UV morphology)
could be playing a role.  We emphasize, however, that the above
arguments are subject to the uncertainties in the scaling factor of
our UV polarized flux, and the low signal to noise of the polarization
measurements.

\subsection{Hidden quasars in other objects}
\label{sec-disc-other}

In the high-redshift narrow-line radio galaxies with extended
scattering regions (over 10kpc scale), 3C256 ($z$=1.824), 3C265
($z$=0.811), and 3C324 ($z$=1.206), their rest-frame UV polarized flux
spectral shape is similar to or rather redder than the typical
quasars. For 3C256 \citep{De96}, the polarized flux spectral index is
$\alpha \simeq -1.1$ ($PF_{\nu} \propto \nu^{\alpha}$) for the rest
wavelength range of $1400-3200$\AA. For 3C265 \citep{Tr98}, $\alpha =
0$ for $\lambda_{\rm rest} = 2200-4900$\AA.  For 3C324 \citep{Ci96},
$\alpha = -2.0 \sim -1.5$ for $\lambda_{\rm rest} = 1800-4100$\AA.
The reddening for the polarized flux in these objects is still
uncertain, though for the total flux of 3C256 the reddening for narrow
lines is inferred to be small from the HeII line ratio \citep{De96},
and for 3C324, \citet{Ci96} concluded that the polarized flux spectrum
is explained by electron or dust scattering only if reddening with
$E(\bv)$ = 0.25 - 0.35 is assumed.

For 3C277.2 ($z$=0.763; Tran et al. 1998), there is slight evidence
for an extended scattering region, and the polarized flux spectrum
shape is roughly $\alpha = -0.6$ at $\lambda_{\rm rest} =
2200-5000$\AA, allowing for a possible 3000\AA\ bump. In 3C356
($z$=1.079; Cimatti et al. 1997) and 4C23.56 ($z$=2.482; Cimatti et
al. 1998b), which are also high-redshift narrow-line radio galaxies,
the morphology is double and one of the two components exhibits high
polarization.  For 3C356, $\lambda_{\rm rest} = 2000-4300$\AA,
\citet{Ci97} found that the polarized flux can be explained either
by electron scattering or by dust scattering + reddening ($E(\bv) =
0.05$). For 4C23.56, $\alpha$ is roughly $-1.5$ for $\lambda_{\rm
rest} = 1200-2600$\AA, and in this far-UV range the polarization
clearly goes up.

In the hyperluminous infrared galaxies, IRAS P09104+4109 ($z$=0.442),
IRAS F15307+3252 ($z$=0.926), and IRAS F10214+4724 ($z$=2.286), the
total flux shows only narrow lines, but the polarized flux reveals
broad permitted lines. Therefore they are interpreted to harbor a QSO,
which is seen in reflection.  For the former two galaxies, the
spectral shape of the polarized flux is again similar to that of a
typical QSO (the continuum shape would not be too different from
quasars; e.g. Cristiani \& Vio 1990) : $\alpha = -0.5$ for IRAS
P09104+4109 at $\lambda\lambda_{\rm rest} 3900-5500$\AA\ (excluding
the 3000\AA\ bump region; Hines et al. 1999; Hines \& Wills 1993), and
$\alpha = -0.5$ for IRAS F15307+3252 at $\lambda\lambda_{\rm rest}
2200-4200$\AA\ \citep{Hi95}. For IRAS F10214+4724, which is a
gravitationally lensed galaxy \citep{Ng99}, the polarized flux is
rather red: $\alpha = -1.4$ at $\lambda\lambda_{\rm rest} 1200 -
2700$\AA\ \citep{Go96}.  For IRAS P09104+4109, the spatial polarized
flux distribution has been found to be extended (a few kpc; Hines et
al. 1999). Also, the extinction of the narrow lines for this galaxy is
inferred to be $A_V=1.8\pm0.6$ \citep{So96}.

\begin{figure*}
\plotone{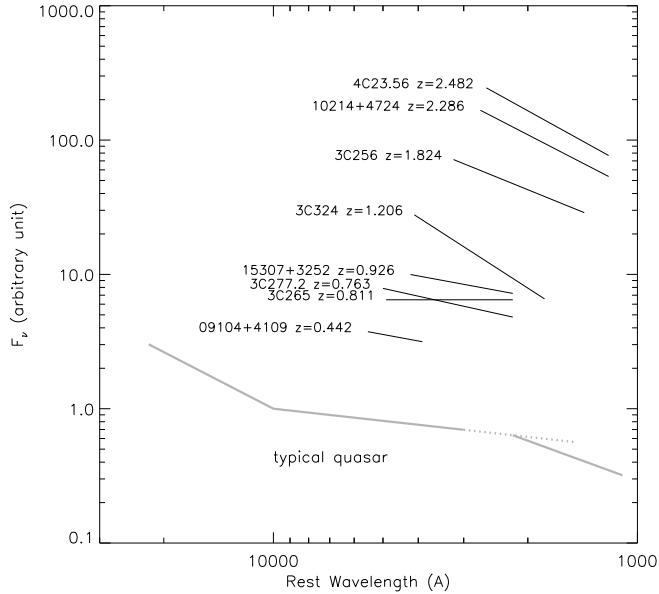}
\figcaption{Illustrates rough spectral shapes of the polarized flux
spectra observed in high redshift galaxies. We have put the spectra in
the order of redshifts. The flux is in $F_{\nu}$, and only the
relative flux in a given object is valid. A typical quasar continuum
shape is also drawn in the thick line at the bottom. We adopted
$\alpha = -0.3$ for the optical spectrum, where $F_{\nu} \propto
\nu^{\alpha}$ \citep{Ne87,Fr91,Fr96}. The spectrum could be steeper,
especially in the UV (so we made the line of $\alpha = -0.3$ dotted in
the UV), as indicated by the quasar composite spectra such as
\citet{CV90} and \citet{Zh97}. The latter far-UV shape is also
illustrated in the thick line at the bottom right. \label{fig-hz}}
\end{figure*}

We have summarized the spectral slopes quoted above in Figure
\ref{fig-hz}. For comparison, the typical quasar spectral shape is
also illustrated (see the first part of this \S\ref{sec-disc} for the
description and references). The comparison seems to be telling us
that we observe the polarized flux spectrum similar to or redder than
the quasar spectrum, but not bluer.  This is why we often cannot rule
out electron scattering even with the poor plausibility. Its
implication is considered in the next section.

\subsection{The case for opaque dust or electron scattering}
\label{sec-disc-sum}

As we have seen, we rarely observe scattered light bluer than the
spectral shape range observed for quasars. Also, the polarized flux
has never been observed to have an exponential cut-off in the shorter
wavelength range so far, which suggests no simple screen
reddening. Therefore, optically-thin dust scattered light would
not probably be dominating the whole scattered flux in general
($\Delta \alpha = + 1$ or more is expected; see
\S\ref{sec-disc-thin}). A possibility is that in a lot of cases, the
significant part of the scattering dust clouds, if there is any, are
probably opaque. In this case, red scattered light without an
exponential decrease in the shorter wavelegth would be naturally
explained.

Is it feasible that the blobs seen in radio galaxies are optically
thick in the UV ?  Here we do some rough estimation using equations
(\ref{eq-ha}) and (\ref{eq-n-f}) in \S\ref{sec-disc-elec}.  Note that
those equations are for optically thin case, but they would be valid
until the optical thickness of the blob becomes unity.  We take
$L_{H\alpha} = 10^{43}$ erg sec$^{-1}$ for a fiducial example.  Then
we obtain $f \sim 10^{-5}$ [eq.(\ref{eq-ha})] assuming the cross
section per H atom = 600$\sigma_T$ for the UV radiation. In this case,
we have $n \sim 500$ cm$^{-3}$ from equation (\ref{eq-n-f}).  On the
other hand, in the HST images of high-redshift radio galaxies
\citep{Be97,Pe99}, the resolved knots in the rest-UV are of $\sim$ 1
kpc scale, and the number of knots in each object is several to $\sim$
10. Therefore, the apparent volume filling factor of these knots in
the whole scattering region ($\sim$ 10 kpc scale or more) is $10^{-2}
\sim 10^{-3}$. This means that the volume filling factor inside the
clump would be about $10^{-5}/10^{-2} \sim 10^{-5}/10^{-3}$. With the
density of the order of $10^2$ cm$^{-3}$, the average column density
through the clump (clump size $\times$ density $\times$ volume filling
factor inside the clump) would be of the order of $10^{21}$
cm$^{-2}$. Thus, this clump will be optically thick in the UV if there
is no particular dust-destroying condition. Therefore, it would be
feasible to have opaque blobs.

If these knots seen in the high-redshift radio galaxies are
proto-galactic subunits as inferred by \citet{Pe99}, these will
certainly provide the actual physical origin of the opaque dusty
clumps discussed here, since such subunits will have a large surface
mass density. Conversely, we can estimate the lower limit for the mass
of an opaque dusty clump from the size scale of the clump (diameter
$2r$) and average column density through the clump ($N_H$) as
\begin{eqnarray} 
M & \simeq & m_H \pi r^2 N_H\\
  & = & 6 \times 10^6 \left( \frac{2r}{1 {\rm kpc}} \right)^2 
        \left( \frac{N_H}{10^{21} {\rm cm}^{-2}} \right) M_{\sun}.
\end{eqnarray}

The possibility for electron scattering, however, still cannot be
ruled out, aside from the large mass problem. The extremely broad
component ($\sim$20000 km sec$^{-1}$ or more), seen in the \ion{C}{4}
line of 3C234 or possibly in the \ha\ lines of 3C234 and 3C321, might
suggest the scattering broadening by hot electrons ($\sim 10^7$ K).
If the spatial distribution of these electrons is really extended to
$\sim$ 10kpc scale and thus have a huge mass, this gas would be
collapsing into the inner region in a timescale shorter than the
dynamical timescale since the cooling time would be of order $10^7$ yr
assuming $n \sim 1$ cm$^{-3}$ (\S\ref{sec-disc-elec}; see Fabian 1989
for the case of larger scale). This would be very important to the
galaxy forming process. The physical origin of this hot gas might be
related to the inner cooled and condensed part of the cooling flow
seen in clusters of galaxies.

\section{Conclusions}\label{sec-conc}

We have presented the results of HST UV spectropolarimetry of
low-redshift narrow-line radio galaxies.  We investigated the
polarized flux spectrum over a wide wavelength range, combining the UV
data with the optical/IR data in the literature.  The polarized flux
spectrum shape in our sample and in other high-redshift radio galaxies
are found to be no bluer than the typical quasar continuum. This, and
apparently no exponential decrease in the shorter wavlength, would
suggest that in a lot of cases the dust scattering clouds, if any,
would be opaque and thus show a grey scattering efficiency. In the
high-redshift galaxies, these opaque dust scattering clouds could be
proto-galactic fragments inferred to be seen in the HST
images. However, the possibility of the scattering by electrons still
cannot be ruled out, which might imply a large gas mass surrounding
these radio galaxies.


\acknowledgments 

The authors acknowledge H. Tran for providing the electronic data for
spectropolarimetry of 3C234.  The authors also thank the referee,
Marshall Cohen, for many useful comments. Support for this work was
provided by NASA through grant number GO-06616 from the Space
Telescope Science Institute, which is operated by AURA, Inc., under
NASA contract NAS5-26555.  This research has made use of the NASA/IPAC
Extragalactic Database (NED) which is operated by the Jet Propulsion
Laboratory, California Institute of Technology, under contract with
the National Aeronautics and Space Administration. M.K. was a Guest
User, Canadian Astronomy Data Centre, which is operated by the
Herzberg Institute of Astrophysics, National Research Council of
Canada.  The work by WvB was performed under the auspices of the
U.S. Department of Energy by University of California Lawrence
Livermore National Laboratory under contract No. W-7405-Eng-48.




\begin{deluxetable}{cccccrrrrrrr}
\tabletypesize{\scriptsize}
\tablecaption{The Data \label{tab-data}}
\tablecolumns{6}
\tablewidth{0pt}
\tablehead{
\colhead{Rootname} & \colhead{Obs Date} &
\colhead{Detector} & \colhead{Grating} & \colhead{Exp Time} & 
\colhead{Mode} 
}
\startdata
\cutinhead{3C234 ($z=0.1848$)}

y3d40104t, y3d40105t  & Dec 8, 1996 & RED  & G190H & $190.0 + 2160.0$&\\
y3d40106t - y3d4010bt & Dec 8, 1996 & BLUE & G270H & 
$2140.0 + 2200.0 \times 5$ & spectropolarimetry\\

\cutinhead{3C321 ($z=0.0961$)}

y3d40204t, y3d40205t  & Aug 25, 1996 & RED  & G190H & $230.0 + 2119.9$&\\
y3d40206t - y3d4020dt & Aug 25, 1996 & BLUE & G270H & 
$2140.0 + 2200.0 \times 7$ & spectropolarimetry\\

\cutinhead{3C327 ($z=0.1048$)}

y3d40304t, y3d40305t  & Aug 10, 1996 & RED  & G190H & $230.0 + 2119.9$&\\
y3d40306t - y3d4030dt & Aug 11, 1996 & BLUE & G270H & 
$2140.0 + 2200.0 \times 7$ & spectropolarimetry\\

\enddata
\end{deluxetable}

\begin{deluxetable}{lcccccccc}
\rotate
\tablecaption{Observed Lines \label{tab-line}}
\tablewidth{0pt}
\tablecolumns{8}
\tablehead{
\colhead{} & \multicolumn{3}{c}{3C234} & \colhead{} &
\multicolumn{3}{c}{3C327}\\
\cline{2-4} \cline{6-8} \\
\colhead{line} &
\colhead{Flux} & \colhead{FWHM} & \colhead{EW} & \colhead{} &
\colhead{Flux} & \colhead{FWHM} & \colhead{EW}
}
\startdata
\ion{Si}{4} $\lambda$1398 + \ion{O}{4}] $\lambda$1402 & 
$45.6\pm7.8$ & $1640\pm310$ & $9.4\pm1.6$ && 
\nodata & \nodata & \nodata\\

[\ion{Si}{7}] $\lambda$1441 &
$10.3\pm3.7$ & $920\pm400$ & $2.2\pm0.8$ &&
\nodata & \nodata & \nodata\\

\ion{N}{4} $\lambda$1487 &
$ 21.1\pm4.0$ & $1050\pm230$ & $4.7\pm0.9$ && 
\nodata & \nodata & \nodata\\

\ion{C}{4}$n$ $\lambda\lambda$1548,1550\tablenotemark{1} & 
$279\pm16$ & $415\pm37$ & $64.7\pm3.8$ && 
$50.5\pm3.9$ & $1080\pm90$ & $96\pm19$ \\

\ion{C}{4}$b$ & 
$140.7\pm9.7$ & $3360\pm250$ & $32.6\pm2.3$ && 
\nodata & \nodata & \nodata\\

\ion{C}{4}$eb$ & 
$206\pm23$ & $20600\pm2100$ & $48.2\pm5.5$ && 
\nodata & \nodata & \nodata\\

\ion{He}{2} $\lambda$1640 & 
$122.5\pm3.3$ & $498\pm16$ & $30.3\pm1.0$ && 
$ 24.1\pm2.0$ & $542\pm54$ & $47.9\pm9.4$ \\

\ion{O}{3}] $\lambda$1661 ? &
$9.3\pm1.9$ & $570\pm160$ & $2.3\pm0.5$ && 
\nodata & \nodata & \nodata\\

\ion{O}{3}] $\lambda$1666 &
$16.3\pm1.9$ & $480\pm76$ & $4.1\pm0.5$ && 
\nodata & \nodata & \nodata\\

\ion{Si}{2} $\lambda$1808 &
$8.4\pm1.0$ & $292\pm53$ & $2.3\pm0.3$ && 
\nodata & \nodata & \nodata\\

\ion{Si}{3}] $\lambda$1892 &
$4.5\pm1.8$ & $830\pm340$ & $1.3\pm0.5$ && 
\nodata & \nodata & \nodata\\

\ion{C}{3}]$n$ $\lambda$1907,1909 & 
$83.2\pm2.3$ & $660\pm20$ & $24.5\pm0.9$ && 
$11.8\pm1.1$ & $525\pm56$ & $20.0\pm2.8$ \\

\ion{C}{3}]$b$ &
$51.5\pm5.8$ & $7030\pm790$ & $15.2\pm1.7$ && 
\nodata & \nodata & \nodata\\

[\ion{Si}{7}] $\lambda$2148 &
$9.3\pm1.1$ & $680\pm100$ & $3.0\pm0.4$ &&
\nodata & \nodata & \nodata\\

\ion{C}{2}] $\lambda$2326 & 
$11.9\pm1.1$ & $858\pm91$ & $3.8\pm0.3$ && 
$ 2.2\pm0.7$ & $800\pm290$ & $4.4\pm1.5$ \\

[\ion{Ne}{4}] $\lambda$2424 & 
$51.4\pm1.3$ & $740\pm20$ & $15.8\pm0.4$ && 
$15.3\pm0.8$ & $748\pm41$ & $26.6\pm1.8$ \\

[\ion{O}{2}] $\lambda$2470 &
$6.3\pm1.2$ & $930\pm190$ & $2.0\pm0.4$ && 
\nodata & \nodata & \nodata\\

\ion{He}{2} $\lambda$2512 + [\ion{Mg}{7}$] \lambda$2509 & 
$6.1\pm1.2$ & $880\pm180$ & $1.9\pm0.4$ && 
\nodata & \nodata & \nodata\\

[\ion{Mg}{7}] $\lambda$2629 & 
$8.3\pm1.0$ & $680\pm100$ & $2.9\pm0.4$ && 
$2.2\pm0.4$ & $580\pm150$ & $4.2\pm0.8$ \\

[\ion{Fe}{11}] $\lambda$2649 ? & 
$4.2\pm0.8$ & $460\pm110$ & $1.5\pm0.3$ && 
\nodata & \nodata & \nodata\\

\ion{He}{2} $\lambda$2733 & 
$5.8\pm0.9$ & $463\pm82$ & $2.0\pm0.3$ && 
\nodata & \nodata & \nodata\\

\ion{Mg}{2}$b$ $\lambda$2796,2805 & 
\nodata & \nodata & \nodata &&
$22.5\pm2.0$ & $8010\pm690$ & $43.7\pm3.9$ \\ 

\enddata

\tablecomments{The mark $n$ is for narrow components, $b$ for broad
components, and $eb$ for extremely broad components. Flux is in units
of $10^{-16}$ erg cm$^{-2}$ sec$^{-1}$.  FWHM is in units of km
sec$^{-1}$. EW is in rest frame and in units of \AA.}

\tablenotetext{1}{For 3C234, the two narrow lines
are fitted as two gaussians and the sum of their fluxes and EWs are
shown.}

\end{deluxetable}

\begin{deluxetable}{cccccccccc}
\rotate
\tabletypesize{\scriptsize}
\tablecaption{Comparison of polarimetric observations. \label{tab-comp}}
\tablewidth{0pt}
\tablecolumns{10}
\tableheadfrac{0.05}
\tablehead{
\colhead{$\lambda_{\rm obs}$} & \colhead{$\lambda_{\rm rest}$} & 
\colhead{Aperture} &
\colhead{$F_{\lambda}$} & \colhead{$P \times F_{\lambda}$} &
\colhead{$F_{\nu}$} & \colhead{$P \times F_{\nu}$} &
\colhead{PA} & \colhead{$P$} & \colhead{Ref} \\
\multicolumn{2}{c}{(\micron)} & \colhead{} &
\multicolumn{2}{c}{(10$^{-17}$ erg cm$^{-2}$ sec$^{-1}$ \AA$^{-1}$)} &
\multicolumn{2}{c}{($\mu$Jy)} & \colhead{($\degr$)} & \colhead{(\%)}
}

\startdata
\cutinhead{3C234} 

$0.22-0.33$ & $0.19-0.28$ & $0.''86$ & $28.6\pm0.1$ & $3.27\pm0.16$ &
$72.5\pm0.2$ & $8.29\pm0.41$ & $143\pm2$ & $11.4\pm0.6$ & 1\\

$0.31\pm0.04$ & 0.26& $4.''64 \times 4.''64$ & 27.8 & $3.34\pm0.14$ &
$89.1$ & $10.7\pm0.4$ & $160 \pm 1$ & $12.0 \pm 0.5$ & 2 \\

$0.44\pm0.05$ ($B$) & 0.37 & $6.''4$ & 38.0 & $3.88\pm0.80$ & 246 &
$25.1\pm5.2$ & $164 \pm 12$ & $10.2 \pm 2.1$ & 3 \\

$0.55\pm0.04$ ($V$) & 0.46 & $9''$ & 54.9 & $2.80\pm0.60$ & 554 &
$28.3\pm6.1$ & $157 \pm 12$ & $5.1 \pm 1.1$ & 3 \\

$0.40-0.90$ & $0.34-0.76$ & $1'' \times 2.''2$ & $21\sim12$ &
$2.6\sim1.3$ & $110\sim320$ & $14\sim35$ & $\sim 157$ & $\sim10$ & 4
\\

$0.61-0.82$ & $0.51-0.69$ & $1''$ slit & $\sim20$ & $\sim2.1$ &
$250\sim450$ & $25\sim50$ & $\sim 154$ & $\sim11$ & 5 \\

$0.64\pm0.07$ ($R$) & $0.54$ & $6''.0$ & 41.8 & $2.67\pm0.56$ & 571 &
$36.5\pm7.7$ & $150\pm6$ & $6.40\pm1.35$ & 6 \\

$1.64\pm0.14$ ($H$) & $1.38$ & $6.''0$ & 15.1 & $0.92\pm0.24$ & 1360 &
$83\pm22$ & $161\pm8$ & $6.08 \pm 1.62$ & 6 \\

$1.97-2.36$ & $1.66-1.99$ & $3.''08 \times 3''$ & $15\sim13$ &
$0.45\sim0.65$ & $2000\sim2400$ & $60\sim120$ & $\sim150$ & $3\sim5$ &
5 \\

$2.2\pm0.2$ ($K$) & $1.9$ & $5.''8~/~7.''5$ & $34\pm4$ & $1.6\pm0.2$ &
$5500\pm600$ & $260\pm30$ & $156\pm2$ & $4.57\pm0.36$ & 7 \\

\cutinhead{3C321}

$0.31\pm0.04$ & 0.28 & $11.''02 \times 11.''02$ & 39.7 & $3.30\pm0.28$
& 127 & $10.6\pm0.9$ & $31 \pm 2$ & $8.3 \pm 0.7$ & 2 \\

$0.31\pm0.04$ & 0.28 & $5''$ & $23.3\pm0.1$ & $1.59\pm0.15$ &
$74.8\pm0.3$ & $5.11\pm0.49$ & $28 \pm 3$ & $6.8 \pm 0.7$ &
1\tablenotemark{a} \\

$0.36\pm0.03$ ($U$) & 0.33 & $5''$ & $34.5\pm0.7$ & $1.41\pm0.16$ &
$149\pm3$ & $6.10\pm0.69$ & $35 \pm 3$ & $4.10 \pm 0.46$ & 8 \\

$0.37-0.83$ & $0.34-0.76$ & $1'' \times 17''$ & $100\sim200$ & $6\sim2$ &
$460\sim4600$ & $28\sim46$  & $30\sim40$ & $6\sim1$ & 9 \\

$0.38-0.63$ & $0.35-0.57$ & $1.''6$ slit & $20\sim40$ & $0.86\sim0.42$ &
$96\sim530$ & $4.1\sim5.6$  & $30\sim40$ & $5\sim1$ & 10 \\

$0.42-0.75$ & $0.38-0.68$ & $2''$ slit & $17\sim48$ & $0.8\sim0.4$ &
$100\sim900$ & $5\sim7$ & $\sim40$ & $1.7\sim0.7$ & 8 \\

$0.43\pm0.05$ ($B$) & 0.40 & $5''$ & $55.5\pm1.2$ & $1.07\pm0.16$ &
$342\pm7$ & $6.6\pm1.0$ & $34 \pm 4$ & $1.93 \pm 0.29$ & 8 \\

$0.54\pm0.04$ ($V$) & 0.50 & $5''$ & $90.3\pm2.0$ & $0.67\pm0.16$ &
$878\pm19$ & $6.5\pm1.6$ & $49 \pm 7$ & $0.74 \pm 0.18$ & 8 \\

$0.65\pm0.07$ ($R$) & 0.58 & $5''$ & $76.0\pm1.7$ & $0.38\pm0.08$ &
$1071\pm24$ & $5.4\pm1.1$ & $37\pm5$ & $0.50\pm0.10$ & 8 \\

$2.2\pm0.2$ ($K$) & 2.0 & $5''$ & $22.3\pm0.5$ & $0.37\pm0.12$ &
$3600\pm81$ & $60\pm19$ & $18\pm10$ & $1.64\pm0.54$ & 8 \\

\cutinhead{3C327}

$0.22-0.33$ & $0.20-0.30$ & $0.''86$ & $5.16\pm0.05$ & $1.03\pm0.09$ &
$13.1\pm0.1$ & $2.62\pm0.23$ & $171\pm3$ & $20.0\pm1.8$ & 1 \\ 

$0.31\pm0.04$ & 0.28 & $5.''30 \times 5.''30$ & 11.7 & $2.35\pm0.29$ &
37.5 & $7.53\pm0.93$ & $160 \pm 4$ & $20.1 \pm 2.5$ & 2 \\

$0.36\pm0.03$ ($U$) & 0.33 & $5''$ & $18.3\pm0.4$ & $1.16\pm0.46$ &
$79.1\pm1.7$ & $5.0\pm2.0$ & $164 \pm 11$ & $6.35\pm2.51$ & 11 \\

$0.43\pm0.05$ ($B$) & 0.39 & $5''$ & $31.1\pm0.7$ & $< 1.0$ &
$192\pm4$ & $< 6.2$ & $125 \pm 72$ & $< 3.2$ & 11 \\

$0.54\pm0.04$ ($V$) & 0.49 & $5''$ & $67.3\pm1.5$ & $< 1.0$ &
$654\pm15$ & $< 9.8$ & $166 \pm 31$ & $< 1.5$ & 11 \\

$0.65\pm0.07$ ($R$) & 0.59 & $5''$ & $59.8\pm1.3$ & $< 0.68$ &
$843\pm18$ & $< 9.6$ & $151 \pm 23$ & $< 1.1$ & 11 \\

$0.79\pm0.07$ ($I$) & 0.72 & $5''$ & $61.5\pm1.3$ & $< 1.6$ &
$1280\pm27$ & $< 34$ & $141 \pm 24$ & $< 2.6$ & 11 \\

$1.64\pm0.14$ ($H$) & 1.48 & $5''$ & $38.7\pm0.8$ & $< 0.74$ &
$3472\pm72$ & $< 67$ & $171 \pm 54$ & $< 1.9$ & 11 \\

\enddata

\tablerefs{1. This paper ; 2. Hurt et al. 1999 ; 3. Cimatti \& di
Serego Alighieri 1995 ; 4. Tran et al. 1995 ; 5. Young et al. 1998 ;
6. Brindle et al. 1990 ; 7. polarization data with $5.''8$ aperture
from Sitko \& Zhu 1991, absolute flux with $7.''5$ aperture from
Lilly, Longair, \& Miller 1985 ; 8. Young et al. 1996a ; 9. Cohen et
al. 2000 ; 10. Tadhunter, Dickson, \& Shaw 1996 ; 11. Young et
al. 1996b }

\tablenotetext{a}{Synthetic $5''$ aperture polarimetry centered on the
SE component, implemented on Hurt et al. (1999) data.}

\end{deluxetable}

\begin{deluxetable}{lcccccc}
\tabletypesize{\scriptsize}
\tablecolumns{4}
\tablecaption{Summary of observed values
\label{tab-scatsize}}
\tablewidth{0pt}
\tablehead{
\colhead{} & \colhead{3C234} & \colhead{3C321} & \colhead{3C327}
}
\startdata
\cutinhead{Size}
size $2R$ $('')$\tablenotemark{a} & 0.5\tablenotemark{a} & 7 & 1 \\
scale for $2R$ (kpc)\tablenotemark{b} & 2 & 17 & 3 \\

\cutinhead{Narrow lines\tablenotemark{c}}
\ha\ flux\tablenotemark{d}
& $5.0\cdot 10^{-14}$  & $4.5\cdot 10^{-14}$ & $2.1\cdot 10^{-14}$\\
\ha\ flux corrected\tablenotemark{e} 
& $8.5\cdot 10^{-14}$  & $1.4\cdot 10^{-13}$ & $5.2\cdot 10^{-14}$\\
\ha\ luminosity\tablenotemark{b,f} 
& $1.4\cdot 10^{43}$   & $5.8\cdot 10^{42}$  & $2.6\cdot 10^{42}$ \\
\ha/\hb  & $3.6\pm0.5$ & $4.4\pm0.6$ & $4.7\pm0.4$ \\
\paa/\ha & $0.18\pm0.02$ & $0.35\pm0.06$ & $0.24\pm0.05$ \\

\cutinhead{Polarized flux}
spectral index $\alpha$\tablenotemark{g}
& $-1.7\pm0.2$ & $-0.1\pm0.3$  & $-0.01\pm0.21$   \\
\ha/\hb & $ 7.3$\tablenotemark{h} & \nodata & \nodata \\

\cutinhead{Broad line in total flux\tablenotemark{c}}
\paa/\ha & $0.52\pm0.11$ & \nodata & \nodata \\

\enddata

\tablenotetext{a}{Approximate FWHM of the polarized flux distribution
from Figs.\ref{fig-pf-a}$\sim$\ref{fig-pf-c}. Note that for 3C234 the
effect of the pre-COSTAR aberration would be large due to the bright
point-like source.}

\tablenotetext{b}{$H_0 = 50$ km sec$^{-1}$ Mpc$^{-1}$ and $q_0=0.5$
are assumed.}

\tablenotetext{c}{The data are from Hill et al. (1996) for $\sim
2.''5$ size.}

\tablenotetext{d}{In units of erg cm$^{-2}$ sec$^{-1}$. }

\tablenotetext{e}{Corrected for the narrow line reddening given in
Hill et al. (1996), in units of erg cm$^{-2}$ sec$^{-1}$.}

\tablenotetext{f}{In units of erg sec$^{-1}$.}

\tablenotetext{g}{For the UV-optical continuum. See \S\ref{sec-disc-ind}
for details}

\tablenotetext{h}{Tran et al. (1995)}

\end{deluxetable}

\end{document}